\documentclass[twocolumn,tighten]{aastex62}
\pdfoutput=1 

\usepackage{amsmath,amssymb,amstext}
\usepackage{apjfonts}
\usepackage[T1]{fontenc}

\newcommand{\avg}[1]{\left\langle{#1}\right\rangle}

\iftrack
\renewcommand{\added}[1]{{\bf #1}}
\renewcommand{\deleted}[1]{}
\renewcommand{\replaced}[2]{{\bf #2}}
\fi

\graphicspath{{./}}

\shorttitle{[\ion{C}{2}] line-intensity mapping prospects}
\shortauthors{Chung et al.}

\begin{document}

\title{Forecasting [\ion{C}{2}] Line-intensity Mapping Measurements Between the End of Reionization and the Epoch of Galaxy Assembly}


\correspondingauthor{Dongwoo T. Chung}\email{dongwooc@stanford.edu}
\author{Dongwoo T. Chung}
\affiliation{Kavli Institute for Particle Astrophysics and Cosmology \& Physics Department, Stanford University, Stanford, CA 94305, USA}
\author{Marco P. Viero}
\affiliation{Kavli Institute for Particle Astrophysics and Cosmology \& Physics Department, Stanford University, Stanford, CA 94305, USA}
\author{Sarah E. Church}
\affiliation{Kavli Institute for Particle Astrophysics and Cosmology \& Physics Department, Stanford University, Stanford, CA 94305, USA}
\author{Risa H. Wechsler}
\affiliation{Kavli Institute for Particle Astrophysics and Cosmology \& Physics Department, Stanford University, Stanford, CA 94305, USA}
\affiliation{SLAC National Accelerator Laboratory, Menlo Park, CA 94025, USA}


\begin{abstract}
We combine recent simulation work on the SFR--[\ion{C}{2}] correlation at high redshift with empirical modeling of the galaxy--halo connection (via \textsc{UniverseMachine}) to forecast [\ion{C}{2}] auto power spectra from $z\sim4$ to $z\sim8$. We compare these to sensitivities realistically expected from various instruments expected to come on-line in the next decade. If the predictions of our model are correct, [\ion{C}{2}] should be detectable up to $z\sim6$ in this generation of surveys, but detecting [\ion{C}{2}] past the end of reionization will require a generational leap in line-intensity survey capabilities.
\end{abstract}




\section{Introduction}
Line-intensity mapping promises unprecedented statistical measurements of high-redshift galaxies---in particular the faint galaxies that dominate luminous activity at high redshift---using emission lines to trace these galaxies in aggregate \citep[for a general overview of the experimental landscape, see][]{Kovetz17}. The [\ion{C}{2}] 157.7 $\mu$m line in particular is a promising choice for its brightness---as bright as 1--2\% of the bolometric far-infrared luminosity of individual low- and high-redshift galaxies---and its role as a tracer of diffuse gas and star-formation activity in the interstellar medium~\citep{Casey14}. Work on this technique is abundant in recent literature, both in signal forecasting \citep{Gong12,Uzgil14,Silva15,Yue15,Serra16,Dumitru18}, and in the design of observational programmes including TIME\footnote{\url{https://cosmology.caltech.edu/projects/TIME}}~\citep{Crites14}, CONCERTO\footnote{\url{https://people.lam.fr/lagache.guilaine/CONCERTO.html}}~\citep{Lagache18IAU}, and CCAT-prime\footnote{\url{http://www.ccatobservatory.org}}~\citep{CCATp}.

All of this work relies to varying degrees on the assumption that [\ion{C}{2}] emission correlates with star-formation rate (SFR), and on an assumption about the form of this relationship. Previous work (in signal forecasting in particular) has relied on either local SFR--[\ion{C}{2}] calibrations (e.g.~\citealt{Spinoglio12} as used in~\citealt{Serra16}; see also \citealt{DeLooze14}, \citealt{HC15} for other local data) or simulations targeting $z>6$ galaxies (e.g.~\citealt{Vallini15}, as used in~\citealt{Yue15}). The recent work of~\cite{Lagache18} connects high-redshift simulations to a comprehensive body of observations down to $z\sim4$, simulating [\ion{C}{2}] galaxies at $z=4$--8 with a modeling approach consistently motivated across the entire redshift range. 
With proposed observations targeting [\ion{C}{2}] emission at various redshifts within a broad range of $z=3$--$14$, a consistently motivated broad- and high-redshift SFR--[\ion{C}{2}] relation, as presented in~\cite{Lagache18}, is necessary for signal forecasting work.

Another component of [\ion{C}{2}] signal forecasting and interpretation, either through analytic halo models or numerical simulations (often only with dark matter), is the galaxy--halo connection (as reviewed in~\citealt{Wechsler18}). This relates the properties of dark matter halos, readily identified in large cosmological simulations, to those of galaxies, readily observed in large\replaced{ }{-}sky surveys. The recent work of~\cite{UniverseMachine} explores such connections using empirical modeling; specifically a forward-modeling framework dubbed~\textsc{UniverseMachine} that uses accretion histories of individual halos with a minimal, flexible galaxy model to track star-formation rates and histories for individual galaxies. The resulting data release includes a catalog of halos with a self-consistent model of the evolution of individual galaxy and host halo properties, with an improved treatment of quenching in massive galaxies compared to previous works. The~\textsc{UniverseMachine} halo catalog thus reflects a particularly \replaced{faithful}{detailed} treatment based on current measurements of the full diversity and stochasticity of galaxy star-formation histories for halos at a given virial mass and redshift, which is necessary when considering emission lines---like [\ion{C}{2}]---tied to star-formation activity.

Here we present signal forecasts for CCAT-prime, CONCERTO, and TIME, and for which we use the \textsc{UniverseMachine} Early Data Release (EDR) of simulated halo catalogues from~\cite{UniverseMachine} with the~\cite{Lagache18} calibration based on the reasoning discussed above. Our simulations specifically explore the expected [\ion{C}{2}] power spectrum and its detectability from the epoch of galaxy assembly ($z\sim3$) to the denouement of the epoch of reionization ($z\sim8$). Note that \cite{Dumitru18} are the first to use the~\cite{Lagache18} calibration for the purpose of [\ion{C}{2}] signal forecasting, but do so only for $z\gtrsim6$; our work covers a more extensive redshift range to encapsulate coverage anticipated from CCAT-prime in particular ($z=3.5$--$8.1$). The paper is structured as follows: in~\autoref{sec:methods} we introduce our methods for simulating [\ion{C}{2}] observations and how they diverge from current observations. We then present the expected signal in~\autoref{sec:results}, 
and present our conclusions in~\autoref{sec:conclusions}.

Where necessary, we assume base-10 logarithms, and the same $\Lambda$CDM cosmology as~\cite{UniverseMachine}: $\Omega_m = 0.307$, $\Omega_\Lambda = 0.693$, $H_0=100h$\,km\,s$^{-1}$\,Mpc$^{-1}$ with $h=0.678$, $\sigma_8 =0.823$, and $n_s =0.96$, all of which should be consistent with the so-called Planck15 cosmology from~\cite{Planck15}. Distances carry an implicit $h^{-1}$ dependence throughout, which propagates through masses (all based on virial masses $\propto h^{-1}$) and volume densities ($\propto h^3$).

\section{Methods}
\label{sec:methods}

\begin{deluxetable*}{cccccccccccc}
\tabletypesize{\footnotesize}
\tablewidth{0.9\linewidth}
\tablecaption{\label{tab:allparams}
Simulation and experimental parameters used for signal and sensitivity forecasts in this work.}
\tablehead{
\colhead{$z$}&\colhead{Frequencies}&\colhead{Lightcone}&\multicolumn{3}{c}{$\sigma_\mathrm{pix}$}&\multicolumn{3}{c}{$N_\mathrm{pix}$}&\multicolumn{3}{c}{$\sigma_\mathrm{pix}t_\mathrm{pix}^{-1/2}$ per beam}\\\cline{4-12}
\colhead{}&\colhead{}&\colhead{size}&\colhead{CCAT-p}&\colhead{CONCERTO}&\colhead{TIME}&\colhead{CCAT-p\tablenotemark{a}}&\colhead{CONCERTO}&\colhead{TIME}&\colhead{CCAT-p}&\colhead{CONCERTO}&\colhead{TIME}\\\cline{4-6}\cline{10-12}
\colhead{}&\colhead{(GHz)}&\colhead{(arcsec)}&\multicolumn{3}{c}{(MJy sr$^{-1}$ s$^{1/2}$)}&\colhead{}&\colhead{}&\colhead{}&\multicolumn{3}{c}{(Jy sr$^{-1}$)}
}
\startdata
3.7&428--388
&180&{2.8}&\nodata&\nodata&1004&\nodata&\nodata&$2.2\times10^4$&\nodata&\nodata\\
4.5&365--325
&165&1.7&18.&\nodata&1004&1500&\nodata&$1.2\times10^4$&$4.7\times10^4$&\nodata\\
6.0&290--250
&150&0.86&11.&11.&1004&3000&32&$6.2\times10^3$&$1.8\times10^4$&$1.6\times10^4$\\
7.4&240--212
&135&{0.70}&7.5&5.2&1004&3000&32&$3.9\times10^3$&$8.0\times10^3$&$5.7\times10^3$
\enddata
\tablenotetext{a}{CCAT-prime instrumental details mean that the number of spatial pixels alone does not inform $t_\mathrm{pix}$ as in the other surveys---see main text (\autoref{sec:sensest}).}
\tablecomments{Lightcone sizes are not field sizes (2 deg$^2$, 1.4 deg$^2$, and $78'\times0.5'$ for CCAT-p, CONCERTO, and TIME respectively), and the latter are used for sensitivity calculations. CCAT-p $\sigma_\mathrm{pix}t_\mathrm{pix}^{-1/2}$ values assume 45$^\circ$ elevation and first-quartile weather with 0.28 mm precipitable water vapour over a 4000-hour survey. For CONCERTO and TIME we use~\autoref{eq:tpix} to calculate $t_\mathrm{pix}$ from $N_\mathrm{pix}$ as well as survey times (1200 hours for CONCERTO and 1000 hours for TIME) and field sizes (mentioned previously), setting $\Omega_\text{pix}=\Omega_\text{beam}$.}
\end{deluxetable*}
\subsection{Experimental Context}
\label{sec:expmeth}

We consider three experiments designed to probe [\ion{C}{2}] at high redshift:
\begin{itemize}
\item The Epoch of Reionization Spectrometer (EoR-Spec) on CCAT-prime (or CCAT-p) is designed for [\ion{C}{2}] line-intensity mapping, covering observing frequencies of $\nu_\mathrm{obs}=210$--420 GHz in two bands, altogether covering $z=3.5$--$8.1$. For this paper, we assume a Phase I instrument with a single dichroic TES bolometer array over 1004 spatial positions occupying one-third of the image plane of one instrument module (of up to seven possible), and a modest resolving power of $R=100$ or a frequency resolution of $\delta_\nu\approx\nu_\mathrm{obs}/100$ throughout the observing bands. The nominal survey programme covers 2 deg$^2$ over 4000 hours.\footnote{Note that the concept for EoR-Spec at first light has evolved during the \replaced{reviewing of this paper; a forecast paper in preparation by the CCAT-prime collaboration will present}{review of this paper; we refer those interested to \cite{Choi19} for} updated first-light EoR-Spec instrumental and survey parameters.}
\item CONCERTO (the CarbON \ion{C}{2} line in post-rEionization and ReionizaTiOn epoch project) is expected to deploy two arrays of spectrometer pixels with channels of $\delta_\nu=1.5$ GHz, with one array observing in 125--300 GHz and the other in 200--360 GHz (or redshift ranges of 5.3--14 and 4.3--8.5). The nominal programme is a survey of 1.4 deg$^2$ over 1200 hours. 
\item TIME (the Tomographic Ionized-carbon Mapping Experiment) is a $R\sim150$ grating spectrometer planning to survey a one-beam-wide $78'\times0.5'$ (or $1.3\times0.0083$ deg$^2$) slice of sky over 1000 hours (\citealt{Crites14}; see also the TIME subsection of~\citealt{Kovetz17}, whose parameters in part supercede that of~\citealt{Crites14}). The spectrometer operates over two bands spanning 183--230 GHz and 230--326 GHz (or redshift ranges spanning 7.3--9.4 and 4.8--7.3). We simplify the resolving power into an optimistic, constant figure of $\delta_\nu=1.5$ GHz for the lower-frequency band and 1.9 GHz for the higher-frequency band.
\end{itemize}

These experiments represent a wide array of state-of-the-art but proven technologies. EoR-Spec on CCAT-p (which we will often refer to simply as CCAT-p in this work) is an evolution of previous spectrometers using Fabry-Perot interferometers (FPI) like SPIFI~\citep{SPIFI,Oberst11}. CONCERTO will use arrays of kinetic inductance detectors (KID) evolved from NIKA~\citep{NIKA} and NIKA2~\citep{Adam18}. TIME inherits the novel waveguide grating spectrometer architecture of Z-Spec~\citep{ZSpec}. Each of these technology sets represents a different approach to enabling a compact, broadband, background-limited, low- to medium-resolution ($R\gtrsim100$) direct-detection spectrometer.

The experiments also represent a range of survey strategies, ranging from TIME's deep line-scan strategy spanning a volume only one beam wide, to CCAT-p's wide-field survey enabled by a degree-scale field of view. Far from being redundant, these three experiments will complement each other in their scope and analysis techniques. However, we are interested in the fundamental ability of each one to achieve a detection of the [\ion{C}{2}] power spectrum.

\subsection{Line Emission Model}
\label{sec:simmeth}
The N-body (dark matter only) cosmological simulation at the base of the \textsc{UniverseMachine} EDR is the Bolshoi-Planck simulation~\citep{BolshoiPlanck1,BolshoiPlanck2}, a periodic box $250h^{-1}$ Mpc on each side with $2048^3$ particles of mass $1.6\times10^8h^{-1}\,M_\odot$. The mass resolution is good enough to resolve a complete sample of halos down to $\sim10^{10}\,M_\odot$; the halo catalogs are incomplete below this mass, which has implications for signal amplitude that we discuss at the end of this section.

We generate sets of 42 lightcones at four different redshift ranges, populated with Bolshoi-Planck dark matter halos with star-formation rates derived from~\textsc{UniverseMachine}, and fully covering the extents outlined in Table~\ref{tab:allparams}. (We generate 58 extra lightcones at the highest redshift of $z\sim7.4$ for a total of 100 lightcones.) CCAT-p has the widest expected spectral coverage of all the instruments we consider, so its coverage influences our choice of redshifts; the CONCERTO and TIME spectral coverages reach the three and two lowest-frequency simulated bands. We use the $250h^{-1}$\,Mpc size of Bolshoi-Planck as an approximate limit for the angular sizes of our lightcones, which are indicated in~\autoref{tab:allparams}. In calculating sensitivities for each experiment, on the other hand, we assume the full expected survey area (but still only the simulated range of frequencies indicated in~\autoref{tab:allparams}).

\begin{figure*}
\centering\includegraphics[width=0.54\linewidth]{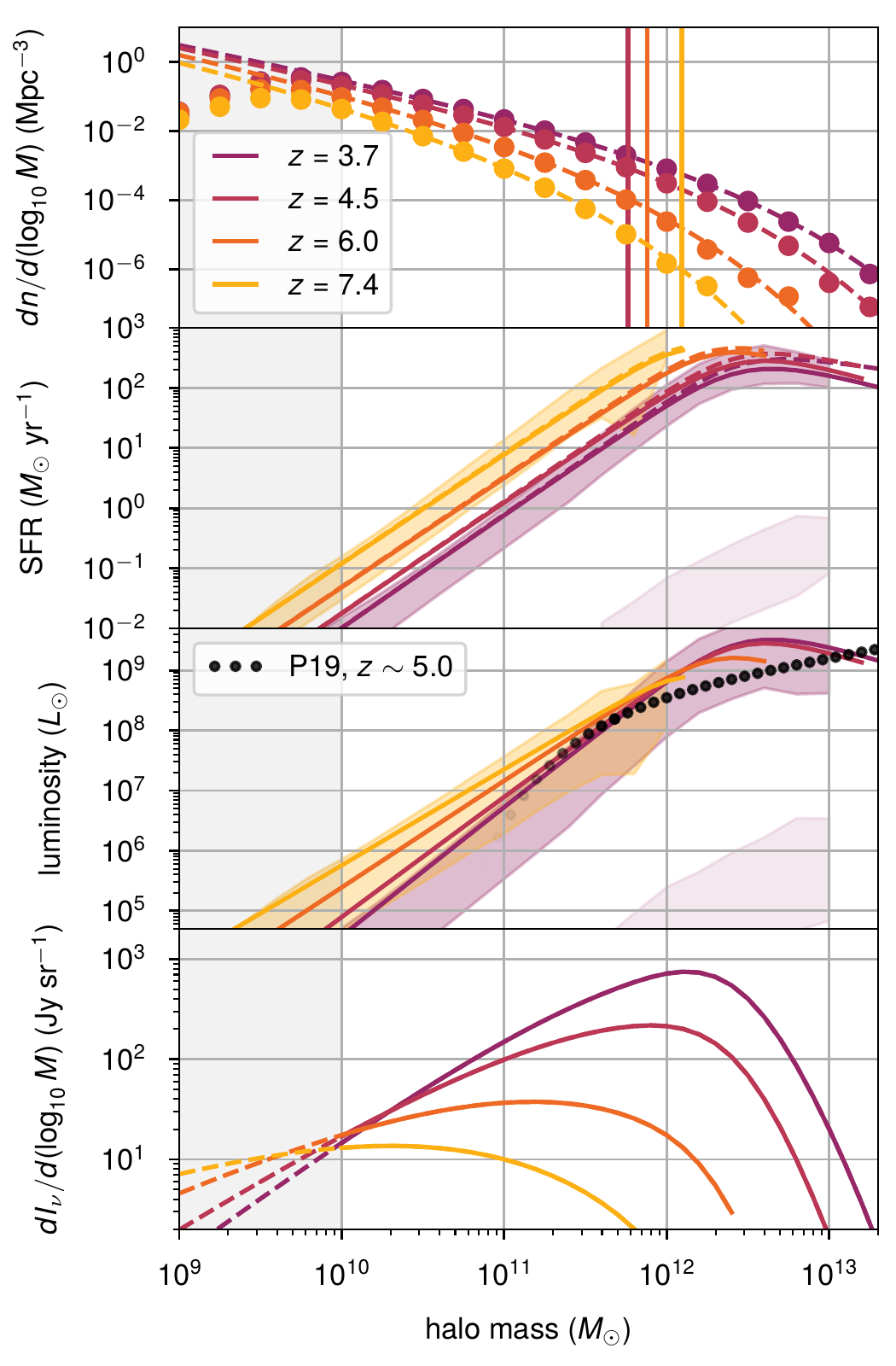}
\caption{\emph{Uppermost panel:} Halo mass function at each indicated redshift in the simulations used in this work. The circles show the values provided in the~\cite{UniverseMachine} \added{early }data release; the dashed curves show an analytic fit described near the start of~\autoref{sec:CIImean}. The vertical lines show the mass above which the quenched fraction for fixed halo mass begins to exceed 10\%. \emph{Upper middle panel:} Mean halo mass--SFR relation expected from~\cite{UniverseMachine} at the indicated redshifts. The solid (dashed) curves show the expected average relation based on the model detailed in~\cite{UniverseMachine}, including (excluding) quenched galaxies; the purple and amber shaded areas show 68\% intervals from simulation snapshots at $z=3.7$ and $z=7.4$, separating star-forming and quenched galaxies. \emph{Lower middle panel:} Expected mean halo mass--[\ion{C}{2}] relation derived from the model detailed in~\autoref{sec:simmeth}, shown at the indicated redshifts. In practice, we derive [\ion{C}{2}] luminosities directly from the star-formation rates calculated for each halo, and again we show sample intervals from simulation snapshots as in the middle upper panel. We also plot the best-fit $z\sim5.0$ halo mass--[\ion{C}{2}] relation from~\cite{Padmanabhan18} (dubbed P1\replaced{8}{9} in the legend), based on local and high-redshift observations (reliable down to $M_\mathrm{vir}\sim10^{11}\,M_\odot$, below which we fade out the plotted curve). \emph{Lowermost panel:} Expected contribution of each halo mass range (calculated in bins of $\Delta(\log{M_\mathrm{vir}})=0.1$, normalised to show the contribution per $\Delta(\log{M_\mathrm{vir}})=1.0$) to mean map intensity, based on the relation in the upper panel and an analytic halo mass function fit as described in~\autoref{sec:CIImean}. Unlike in the previous panels, we show averages over the redshift ranges used in our simulations, indicated in~\autoref{tab:allparams}. In all panels, we shade the part of the axes corresponding to $M_\mathrm{vir}<10^{10}\,M_\odot$, a mass range that is not reflected at all in the $P(k)$ simulations of this work (but is in e.g.~\citealt{Dumitru18}).}
\label{fig:CII_analytic}
\end{figure*}

We assign each halo in the lightcones a luminosity based solely on its star-formation rate\footnote{The \textsc{UniverseMachine} EDR assigns an `observed' SFR to each halo in addition to the `true' SFR, with the former accounting for common observational systematics that result in inaccurate recovery of the latter in real-world data. \cite{Lagache18} largely take SFR values from their semi-analytic model at face value for their SFR--[\ion{C}{2}] relation; they consider observed SFR values incorporating UV attenuation at one point but this does not appear to result in any systematic offset. Therefore, we use the `true' SFR instead of the `observed' SFR in this work.} and cosmological redshift (we ignore peculiar velocities and redshift-space distortions throughout). We impose a minimum emitter halo mass of $10^{10}\,M_\odot$, and for halos above this mass, we use a power-law SFR--[\ion{C}{2}] luminosity relation:
\begin{equation}
\log{\left(\frac{L_\mathrm{line}}{L_\odot}\right)} = \alpha\log{\left(\frac{\mathrm{SFR}}{M_\odot\,\mathrm{yr}^{-1}}\right)} + \beta.
\label{eq:CIIcal}
\end{equation}
\cite{Lagache18} find a redshift-dependent mean relation with $\alpha=1.4-0.07z$ and $\beta=7.1 - 0.07z$. We assume 0.5 dex scatter on the SFR--[\ion{C}{2}] relation, based on the 0.5--0.6 dex dispersion that~\cite{Lagache18} find. The middle panels of~\autoref{fig:CII_analytic} illustrate the mean halo mass--SFR relation and the mean halo mass--[\ion{C}{2}] luminosity relation at the simulated redshifts, including the quenched fractions prescribed by~\cite{UniverseMachine}\footnote{A close inspection of~\autoref{fig:CII_analytic} shows that quenching is not the sole reason for the luminosity ultimately declining with higher halo mass, although it contributes significantly. The~\cite{UniverseMachine} model assumes not only that the star-formation efficiency peaks at a certain halo mass, but is actually boosted around this mass, so the SFR and thus [\ion{C}{2}] luminosity declines slightly as the boost disappears.}. As the figure shows, our mean relation compares favorably to the best-fit model from~\cite{Padmanabhan18}, which comes from relying on abundance matching at $z\sim0$ against the~\cite{Hemmati17} luminosity function and inferring redshift evolution based on constraints from~\cite{Pullen18} on integrated [\ion{C}{2}] emission at $z\sim2.6$. Note that~\cite{Padmanabhan18} finds that current observations only allow constraints down to $M_\text{vir}=10^{11}\,M_\odot$, and~\cite{Lagache18} state that their average relation describes [\ion{C}{2}] luminosities down to $10^7\,L_\odot$.

Note again that the mean SFR--[\ion{C}{2}] relation in~\cite{Lagache18} derives from simulations of high-redshift galaxies using semi-analytic modeling (\texttt{G.A.S.}; updated from~\citealt{Cousin15a,Cousin15b,Cousin16}) and a photoionization code (\texttt{CLOUDY};~\citealt{CLOUDY}), rather than observations of local galaxy samples as in~\cite{Spinoglio12} (a synthesis of data from~\citealt{Brauher08}),~\cite{DeLooze14} and~\cite{HC15}. As~\cite{Lagache18} note, analysis of line observations plus heating and attenuation effects from the cosmic microwave background suggest that high-$z$ [\ion{C}{2}] emission is dominated by ionized carbon in photo-dominated regions (PDR), rather than neutral gas as appears to be the case in local star-forming galaxies. By combining~\texttt{G.A.S.} modeling of galaxy formation history with~\texttt{CLOUDY} modeling of the PDR in each galaxy,~\cite{Lagache18} should represent a state-of-the-art understanding of [\ion{C}{2}] emission at high redshift.

At $z\sim6$, the~\cite{Lagache18} SFR--[\ion{C}{2}] relation takes $\alpha=0.98$ and $\beta=6.68$. The local calibrations by comparison would assign luminosities 2--10 times higher at a given SFR. For~\cite{HC15}, $\alpha=0.967$ and $\beta=7.65$; for~\cite{DeLooze14}, $\alpha=0.99$ and $\beta=6.92$ (based on the entire literature sample); and $\alpha=0.89$ and $\beta=7.27$ for~\cite{Spinoglio12} (converting the original $L_\text{IR}$--[\ion{C}{2}] relation to a SFR--[\ion{C}{2}] one by taking $L_\text{IR}/L_\odot=10^{10}\,\operatorname{SFR}/(M_\odot\text{ yr}^{-1})$)\footnote{The relation used in \cite{Serra16} does not take into account the erratum issued for~\cite{Spinoglio12} correcting IR luminosities up by a factor of $1.8$~\citep{Spinoglio12E}; the resulting change in the inferred $L_\text{IR}$--[\ion{C}{2}] relation would move $\beta$ down to 7.04.}.

We use \texttt{limlam\_mocker}\footnote{\url{https://github.com/georgestein/limlam_mocker/}} to generate line-intensity cubes and power spectra for each lightcone using this model. Doing this requires defining a grid of volume elements \added{(}or voxels\added{)}, each taking up a solid angle and frequency interval within the mocked line-intensity cube. 
We use the frequencies and lightcone sizes in~\autoref{tab:allparams} and create an intensity cube of $450^3$ voxels\footnote{None of the three surveys could produce a grid of $450^3$ well-resolved voxels with the specified angular sizes. However, this part of the work requires accurate forecasting of the signal more than faithful mocking of observations for each survey. We consider survey limitations in~\autoref{sec:sensest}.}. 
All halo luminosities are binned per voxel, and the [\ion{C}{2}] luminosity per voxel $L_\text{vox}$ is converted into an intensity $I_\nu=L_\text{vox}/(4\pi D_L^2\Omega_\text{pix}\delta_\nu)$ by dividing by the voxel frequency interval $\delta_\nu$, voxel solid angle $\Omega_\text{pix}$, and $4\pi D_L^2$ given the luminosity distance $D_L$ from the observer to the voxel. We can then calculate both the mean map intensity $\avg{I_\nu}$ from the simulated intensity cube, presented in~\autoref{sec:CIImean}, and the spherically averaged, comoving 3D power spectrum $P(k)$, which we present as the main signal in~\autoref{sec:pspec}. The latter is obtained from the full 3D power spectrum $P(\mathbf{k})$ of the intensity cube in comoving space, averaged in spherical $k$-shells.

\added{Note that while $\avg{I_\nu}$ is not to be measured directly by any of the experiments we consider, it is nonetheless a useful and potentially important statistic. Unlike $P(k)$, it can be calculated analytically purely from the halo mass function with the model outlined above. Thus, analytic expectations for $\avg{I_\nu}$ will act as a sanity check on our simulation results.}

Given that analysis of recent observations points to star formation arising in $z\gtrsim6$ galaxies with halo masses as low as several $10^9\,M_\odot$~\citep{Finlator17}, the minimum emitter halo mass of $10^{10}\,M_\odot$ imposed in these simulations is potentially too high, but is forced by the Bolshoi-Planck mass resolution, which results in halo incompleteness below $\sim10^{10}\,M_\odot$. \replaced{Thus $\avg{I_\nu}$ is a useful and potentially important statistic, as it can be calculated analytically purely from the halo mass function with the model outlined above, which is not the case for $P(k)$.}{However, analytic calculations of $\avg{I_\nu}$ are not subject to the same limits.} To gauge the effects of the halo mass function and the minimum emitter halo mass\added{ on the signal}, we make analytic estimates of $\avg{I_\nu}$ at each redshift using the above model with a halo mass function (HMF) fit at each redshift that should predict the correct abundances of halos with masses $\lesssim10^{10}\,M_\odot$. In the process, we also calculate the expected contribution to the mean intensity from different mass bins. We present these results alongside the simulation results (for total $\avg{I_\nu}$ only) in~\autoref{sec:CIImean}.

\subsection{Sensitivity Estimates}
\label{sec:sensest}

We follow the formalism of~\cite{Li15} and quantify the uncertainty in $P(k)$ as
\begin{equation}
\sigma_P(k) = \frac{P(k)+P_n}{\sqrt{N_m(k)}},
\label{eq:totalsigma}\end{equation}
which is the total observed power spectrum---signal $P(k)$ plus noise $P_n$---divided by the number of Fourier modes $N_m(k)$ available for averaging near that given $k$. We detail the calculation of $N_m(k)$ in~\autoref{sec:Nmodes}.

We calculate $P_n$ from instrumental noise only. If our survey volume is observed uniformly so that each volume element (or voxel) of some comoving volume $V_\mathrm{vox}$ has been observed for some integration time $t_\mathrm{pix}$, then
\begin{equation}P_n=\frac{\sigma_\mathrm{pix}^2}{t_\mathrm{pix}}V_\mathrm{vox},\end{equation}
where $\sigma_\mathrm{pix}$ is the sensitivity per instrumental pixel per spectral element, and $\sigma_\mathrm{pix}t_\mathrm{pix}^{-1/2}$ the final survey sensitivity per voxel (the equivalent of $\sigma_n$ in Appendix C of~\citealt{Li15}). We show $\sigma_\mathrm{pix}$ in~\autoref{tab:allparams} for each survey, quantified as noise-equivalent intensity/input (NEI), as well as the number of instrumental pixels expected (per band). We explain the figures for each survey in more detail below.
\begin{itemize}
\item For CCAT-p, we use figures for $\sigma_\mathrm{pix}t_\mathrm{pix}^{-1/2}$ per beam for a single EoR-Spec array given by the collaboration (G.~Stacey, private communication).\added{ The figures are calculated at the central frequencies of each simulated frequency band, assuming first-quartile weather (0.4 mm precipitable water vapour) and an observing elevation of 45 degrees.} Each array will have 1004 spatial beams sensitive to two polarisations (which are folded into the $\sigma_\mathrm{pix}$ given), but instrumental details mean that only a fraction of the total spectral coverage can be instantaneously observed and the FPI must step across multiple settings to sample the full bandwidth, which slightly complicates the calculation of $t_\mathrm{pix}$. However, broadly speaking, the results are equivalent to taking $N_\mathrm{pix}$ to be 1004 spatial beams times a factor $\ll1$ which depends on the observing frequency, for an effective count of around 20.
\item CONCERTO will have an array of 1500 pixels for each band, for a total of 3000 pixels in the 200--300 GHz overlap between the two bands (G.~Lagache, private communication). The overlap excludes $z=4.5$ but includes our two highest simulated redshifts. We use the noise-equivalent flux density (NEFD) given in~\cite{Serra16} of 155 mJy s$^{1/2}$ and divide by the beam solid angle to obtain the NEI.\replaced{ (Note}{

We wish to make clear a few caveats around this NEFD value. The most important caveat is that only one value is given with no frequency dependence. In reality, atmospheric opacity is far greater at 325--365 GHz compared to below 300 GHz, so we expect higher NEFD near 345 GHz versus at lower frequencies. Another caveat worth noting is} that \cite{Serra16} actually consider the NEFD divided by $\sqrt{N_\text{pix}}$; however, this is still used as if it were merely the NEFD per pixel, since their expression for $t_\text{pix}$ also includes a factor of $N_\text{pix}$. That said, we show in~\autoref{sec:CCAT-psens} that the 155 mJy s$^{1/2}$ figure is in principle not unreasonable as a NEFD per pixel for CONCERTO. Note\replaced{ also}{, though,} that while \cite{Dumitru18} use the same NEFD while referring to the NIKA2 sensitivities demonstrated in~\cite{Adam18}, \replaced{the NIKA2 sensitivities cannot be extrapolated to this NEFD per pixel without significant assumptions about improvements in system efficiency and emissivity.)}{our calculations in~\autoref{sec:CCAT-psens} make far more optimistic assumptions about system efficiency and emissivity than NIKA2 operating conditions would indicate. (We also still end up with somewhat higher estimated noise, but within a factor of order unity of the figures provided by the CONCERTO team.)

Ultimately, for the purposes of this work, we will use the 155 mJy s$^{1/2}$ figure as is at all frequencies, and we leave a re-examination of CONCERTO sensitivities for future work from the CONCERTO team or others.}
\item For TIME, we use the median of the \added{NEI }range\added{s for the low- and high-frequency bands} quoted for operation at the Arizona Radio Observatory (ARO) assuming 3 mm precipitable water vapour (TIME Collaboration, private communication). \cite{Crites14} also indicate that the TIME experiment will have 32 spectrometers (16 per polarisation).
\end{itemize}

Converting between NEFD and NEI requires knowledge of the beam width, quantified as the full width at half maximum (FWHM). Since TIME and CONCERTO are both to operate on 12-metre telescopes, we assume that the beams for both instruments have a diffraction-limited FWHM of $1.22\lambda/(12\text{ m})$, ranging from $17''$ to $31''$ throughout the full spectral range. (\citealt{Sun16} and~\citealt{Dumitru18} assume similarly for TIME and CONCERTO.) For the CCAT-p beam FWHM, we use figures provided by the collaboration of $(37'',39'',46'',53'')$ at $(408,345,280,214)$ GHz.

For CONCERTO and TIME, which should have simultaneous uniform coverage of all frequency channels within each band by virtue of their architectures, we take $t_\mathrm{pix}$ to be simply the integration time per pixel:
\begin{equation}t_\mathrm{pix}=\frac{N_\mathrm{pix}t_\mathrm{surv}}{\Omega_\mathrm{surv}/\Omega_\mathrm{pix}},\label{eq:tpix}\end{equation}
which is to say the total survey time multiplied by the number of instrumental pixels, divided by the number of map pixels (the ratio of the survey solid angle to the pixel solid angle). For CCAT-p, as noted above, the instantaneous spectral coverage and thus the calculation of $t_\mathrm{pix}$ is more complex, and $\sigma_\mathrm{pix}t_\mathrm{pix}^{-1/2}$ is presented per beam. We present the same in~\autoref{tab:allparams} for CONCERTO and TIME, using $\Omega_\text{pix}=\Omega_\text{beam}$.

The only input left for $P_n$ is the comoving volume per voxel. Since $V_\mathrm{vox}\propto\Omega_\mathrm{pix}\delta_\nu$, this cancels out the $\Omega_\mathrm{pix}$ dependence of $t_\mathrm{pix}$ and the $\delta_\nu^{-1/2}$ dependence of $\sigma_\text{pix}$ (see~\autoref{sec:CCAT-psens} for a detailed explanation) when calculating $P_n$, which thus does not depend on the voxel extent in any dimension. The $\sigma_\mathrm{pix}t_\mathrm{pix}^{-1/2}$ obtained for all experiments is per beam, so we take $V_\mathrm{vox}$ to be the comoving volume within a solid angle of $\Omega_\mathrm{beam}$ and a frequency interval of $\delta_\nu$.

Having calculated $P_n$ and thus $\sigma_P(k)$ for each experiment, we must finally consider attenuation of the observed power spectrum at high wavenumber $k$ due to the finite beam size of each telescope. This attenuation $W(k)$ of the signal results in an effective sensitivity limit of $\sigma_P(k)/W(k)$. Appendix C.3 of~\cite{Li15} details an analytic calculation of $W(k)$, but we make an analogous numerical calculation in this work based on the expected voxel grid for each experiment. For this calculation only, we assume map pixel widths of $15''$ for CCAT-p and $5''$ for the other experiments. Once the pixel width is finer than the standard deviation of the Gaussian beam profile (FWHM$/2.355$), the degree of angular oversampling makes little difference in the numerical calculations and resulting $W(k)$.

\section{Results and discussion}
\label{sec:results}
\subsection{Mean map intensity}
\label{sec:CIImean}

\autoref{tab:CIImean} shows the mean map intensity both from our simulated survey volumes and from the expected contributions from halos based on analytic calculations. The latter uses the best-fit model from~\cite{UniverseMachine} with the HMF of~\cite{SMT}, modified to fit the HMF values provided in the~\cite{UniverseMachine} EDR at each redshift. For the HMF, setting $A=0.62-0.071z+0.0039z^2$ and $a=0.96-0.072z+0.0058z^2$---instead of the original redshift-independent values of $A\approx0.322$ and $a=0.707$ from~\cite{SMT}---provides an adequate description of the actual HMF down to $M_\text{vir}=10^{10}\,M_\odot$ at the redshifts considered here (as we illustrate in the uppermost panel of~\autoref{fig:CII_analytic}). Since the best-fit model prescribes average star-formation rates and quenched fractions as functions of the halo maximum circular velocity at peak historical virial mass $v_{M_\mathrm{peak}}$, rather than of virial mass, we use the relation given in Appendix E2 of~\cite{UniverseMachine} to convert the model relations into functions of virial mass\footnote{We do modify the exponent in Equation E2 from 3 to 0.3, which is necessary for a reasonable approximation to the halo mass--SFR relation observed in the simulation.\added{ This also conforms better to the virial expectation of $v\sim M^{1/3}$.}}. This mass--$v_{M_\mathrm{peak}}$ relation is inexact (with~\citealt{UniverseMachine} indicating scatter of $\sim0.1$ dex), meaning $\sim10\%$ discrepancies between our numerical results and analytic results for $\avg{I_\nu}$ should not be surprising. Indeed, any discrepancies between analytic and simulated results in~\autoref{tab:CIImean} are within this expectation.

\begin{deluxetable}{cccc}
\tabletypesize{\footnotesize}
\tablewidth{0.9\linewidth}
\tablecaption{\label{tab:CIImean}Mean map intensities calculated through both simulations (median and 90\% sample interval) and an analytic halo mass function (HMF) fit with different minimum emitting halo masses.}
\tablehead{
\colhead{Redshift}&\multicolumn{3}{c}{$\avg{I_\nu}$ (Jy sr$^{-1}$)}\\\cline{2-4}
\colhead{}&\colhead{90\% interval}&\multicolumn{2}{c}{Analytic, $M_\mathrm{vir,min}=\cdots$}\\\cline{3-4}
\colhead{}&\colhead{from simulations}&\colhead{$10^{10}\,M_\odot$}&\colhead{$10^{9}\,M_\odot$}}
\startdata
3.7&$924.2_{-54.1}^{+35.1}$&865.0&870.3\\[2pt]
4.5&$339.2_{-9.7}^{+14.5}$&308.4&315.6\\[2pt]
6.0&$64.74_{-2.33}^{+3.08}$&63.48&73.42\\[2pt]
7.4&$16.58_{-0.69}^{+0.66}$&17.55&27.81\\[2pt]
\enddata
\tablecomments{The analytic calculation uses an approximation to the~\cite{UniverseMachine} mean halo mass--SFR relation, and includes no scatter in SFR or line luminosity. The analytic HMF used is a modification of the fit in~\cite{SMT} described in the main text.} 
\end{deluxetable}

The lowermost panel of~\autoref{fig:CII_analytic} shows the expected relative contribution of different halo mass ranges to $\avg{I_\nu}$. At $z=5.8$ and below, the slope of the halo mass--[\ion{C}{2}] relation is steep enough compared to the slope of the HMF that a majority of $\avg{I_\nu}$ comes from halos of masses $\gtrsim10^{11}\,M_\odot$. Lowering the minimum emitting halo mass to $10^9\,M_\odot$ (with a simple extrapolation below $10^{10}\,M_\odot$) thus has little effect on the analytically estimated $\avg{I_\nu}$ below $z\sim6$, increasing only by 0.6\% at $z=3.7$, 2.3\% at $z=4.5$, and 15.6\% at $z=6.0$.

By $z=7.4$, however, halos with $M_\text{vir}\lesssim10^{11}\,M_\odot$ contribute a majority of the average signal, and lowering the cut to $10^9\,M_\odot$ would allow more low-mass halos to contribute, increasing the analytically calculated mean [\ion{C}{2}] intensity by a factor of 1.6. This roughly translates to an increase in the power spectrum ($P(k)\sim\avg{I_\nu}^2$) by a factor of 3. The analytic $\avg{I_\nu}$ with $M_\mathrm{vir,min}=10^{10}\,M_\odot$ is within 10\% of the numerically simulated mean intensity, so our $P(k)$ forecast may indeed be too low by a factor of 3 due to an overly high cutoff mass.

Whether lowering the cutoff halo mass below $10^{10}\,M_\odot$ is well-motivated is unclear. As noted above, \cite{Finlator17} does find evidence for unsuppressed star-formation activity in $z\gtrsim6$ halos with masses below $10^{10}\,M_\odot$, but the evidence is not strong at $z=8$ and not very strong at $z=7$, and $z\sim7.4$ is the only redshift at which a lower cutoff significantly affects $\avg{I_\nu}$ in~\autoref{tab:CIImean}. We discuss the cutoff further in the next section, although we find that even a factor-of-3 increase in the signal at $z\sim7.4$ will require greatly upgraded surveys to distinguish or detect.

\subsection{Power spectra and comparison to previous work}
\label{sec:pspec}

\begin{figure*}
\centering\includegraphics[width=0.72\linewidth]{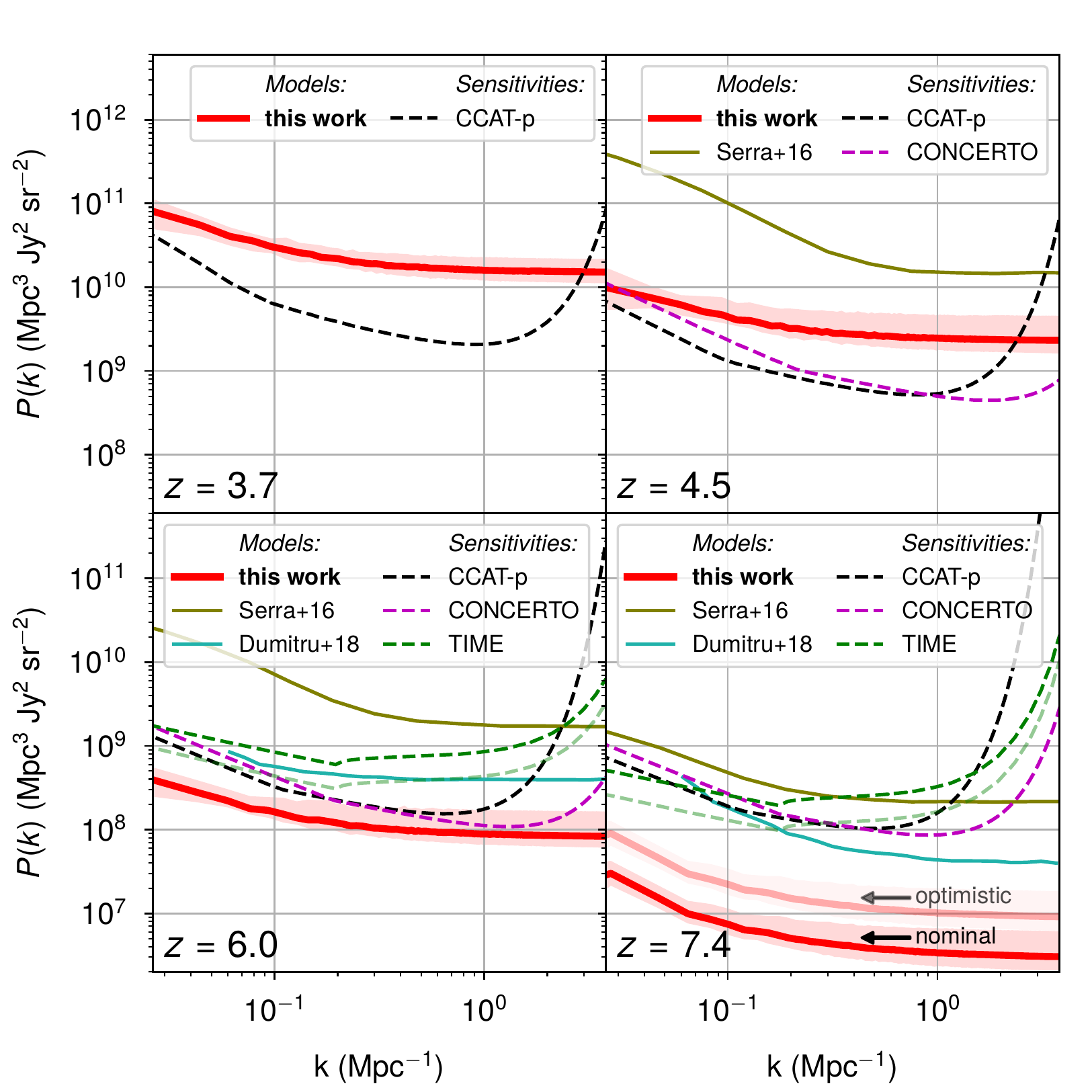}
\caption{Median and 90\% sample interval $P(k)$ values (red solid curves and shaded areas) from simulations at four different redshifts (shown at lower left corner of each panel), and expected $1\sigma$ sensitivity limits (dashed curves) for CCAT-p, CONCERTO, and TIME (black, magenta, and green) given $k$-bins of width $\Delta k=0.035$ Mpc$^{-1}$. We also show a crude estimate of the sensitivities expected from TIME at the Chajnantor plateau instead of at ARO (faint green)---with most other survey and instrumental parameters kept the same---which should halve the noise power spectrum $P_n$. The survey bandwidth assumed is $\Delta\nu=40$ GHz except at $z=7.4$, where $\Delta\nu = 28$ GHz. We plot $\sigma_P(k)/W(k)$ instead of just $\sigma_P(k)$ to show signal-to-noise attenuation due to beam size. At $z=7.4$ only, we show an `optimistic' forecast (faint red solid curve and shaded area) above the \replaced{fiducial one}{nominal fiducial forecast}, emulating a lower minimum [\ion{C}{2}] emitter halo mass than simulations allow. We also show $P(k)$ from~\cite{Serra16} and~\cite{Dumitru18} interpolated or extrapolated to the indicated redshift (other solid curves).}
\label{fig:pspec}
\end{figure*}
We show the $P(k)$ values calculated from the simulations in~\autoref{fig:pspec}. Our forecast signal level appears to drop an order of magnitude or so with each increase in observed redshift. However, in light of the analytic checks of~\autoref{sec:CIImean}, we consider the possibility that the decline in $P(k)$ between $z=6.0$ and $z=7.4$ depends on the choice of cutoff halo mass. If halos below our chosen cutoff of $10^{10}\,M_\odot$ emit in [\ion{C}{2}], our simulated $P(k)$ at $z=7.4$ may be an underestimate by a factor of several. Therefore, we also show an `optimistic' forecast at $z=7.4$ alongside the fiducial one in~\autoref{fig:pspec} by multiplying the fiducial $P(k)$ by a factor of 3 (suggested by the comparison to analytic calculations in~\autoref{tab:CIImean}).

We put our predictions in the context of previous work by plotting them together with $P(k)$ from~\cite{Serra16} (which claims agreement with~\citealt{Gong12}) and~\cite{Dumitru18}\footnote{Throughout this work, we use results provided by Dumitru (private communication) which differ from the initial preprint but should be reflected in the final published work of~\cite{Dumitru18}. In particular, the $P(k)$ values have been revised slightly downwards from the initial preprint, and the cosmic SFR density revised slightly upwards (particularly near $z=6$).}, interpolated or extrapolated in redshift as necessary (as the approximate evolution of $P(k)$ with redshift at each $k$ is apparent in both works). \cite{Serra16} use measurements of the cosmic infrared background anisotropies with a halo model to constrain a halo mass--infrared luminosity relation, and combine this with the local $L_\text{IR}$--[\ion{C}{2}] relation of~\cite{Spinoglio12} to relate halo mass to [\ion{C}{2}] luminosity and thus enable an analytic calculation of $P(k)$. Given our discussion about the difference between local SFR--[\ion{C}{2}] calibrations and the relation from~\cite{Lagache18} being as large as 1 dex at $z\sim6$, we find it unsurprising that the $P(k)$ values of~\cite{Serra16} are almost 2 dex higher than our predictions.

The work of~\cite{Dumitru18} is more similar to ours in comparison, assigning [\ion{C}{2}] luminosities to halos identified in cosmological simulations to directly simulate cubes of [\ion{C}{2}] intensity, and using~\cite{Lagache18} for the SFR--[\ion{C}{2}] scaling relation in their model. Given these similarities, the discrepancy between our prediction and theirs is more surprising at first glance. However, the model of~\cite{Dumitru18} prescribes star-formation rates that are simply proportional to halo mass, does not model a quiescent galaxy population, and results in a somewhat higher cosmic star-formation rate density at $z\sim6$ compared to \textsc{UniverseMachine}. These differences between the halo mass--SFR relations of the two models help explain the factor-of-4 difference we see between our forecast $P(k)$ and the extrapolation from the results of~\cite{Dumitru18} at $z\sim6$.

At $z\sim7$--8, the two models result in a more similar cosmic SFR density, but still prescribe SFR in halos in substantively different ways. Additionally, the minimum halo mass of $2.3\times10^8h^{-1}\,M_\odot=3.4\times10^8\,M_\odot$ in the simulations that~\cite{Dumitru18} use (with no additional cutoff imposed for [\ion{C}{2}] emission) becomes a more significant source of discrepancy at these highest redshifts. We have stated above that a lower cutoff halo mass in our simulations could increase our forecast $P(k)$ by a factor of 3, but this is merely a zeroth-order estimate. A perfectly fair comparison at these highest redshifts against the results of~\cite{Dumitru18} would require deploying the \textsc{UniverseMachine} framework on a simulation fine enough to allow resolution of halos with $M_\text{vir}=10^9\,M_\odot$. This would enable [\ion{C}{2}] simulations that incorporates a more complete halo population, although it is an entirely open question as to how well-justified it would be either to set a lower cutoff mass or to extrapolate the $M_\text{vir}$--[\ion{C}{2}] relation so far below $10^{11}\,M_\odot$ in halo mass or $10^7\,L_\odot$ in line luminosity. We leave this for possible future work.

Finally, while we do not explicitly plot $P(k)$ from~\cite{Padmanabhan18} to compare, there is broad agreement here with that work, with our $P(k)$ below the best-fit model but still within the associated uncertainties. This is to be expected given the level of agreement in the [\ion{C}{2}] luminosity prescription between our model and that of~\cite{Padmanabhan18} already shown in~\autoref{fig:CII_analytic}.

\subsection{Detectability of power spectra}
\label{sec:SNR}

Between our work and previous work considered in~\autoref{sec:pspec}, predictions for the [\ion{C}{2}] signal span a range of several orders of magnitude, unconstrained by any observational data. An improved understanding of [\ion{C}{2}] emission and its connection to star-formation activity at high redshift would be made possible with a $P(k)$ detection or even an upper limit that could exclude the more optimistic models. To consider the ability of near-future surveys to do this, we return to the sensitivities considered in~\autoref{sec:sensest}.

\begin{deluxetable*}{rccccccccc}
\tabletypesize{\footnotesize}
\tablewidth{1.618\linewidth}
\tablecaption{\label{tab:totSNR}
Total signal-to-noise ratio summed over all scales up to $k=1$ Mpc$^{-1}$, and number of hours required to obtain a signal-to-noise ratio of 1 in a $k$-bin centred at $k=0.026$ Mpc$^{-1}$ of width $\Delta k=0.035$ Mpc$^{-1}$, for all experiments considered in this work. We show the mean all-$k$ $\mathrm{S/N}$ and median required survey time (rounded up to two significant figures) across all lightcones.}
\tablehead{\colhead{Experiment}&\colhead{Nominal}&\multicolumn{4}{c}{$\mathrm{S/N}$ across all $k$}&\multicolumn{4}{c}{Survey time required for $\mathrm{S/N}=1$}\\[-1em]\colhead{}&\colhead{survey time}&\multicolumn{4}{c}{}&\multicolumn{4}{c}{at $k=0.026\pm0.0175$ Mpc$^{-1}$}\\[-0.5em]\colhead{}&\colhead{(hours)}&\multicolumn{4}{c}{}&\multicolumn{4}{c}{(hours)}\\\colhead{}&\colhead{}&\colhead{$z=3.7$}&\colhead{$z=4.5$}&\colhead{$z=6.0$}&\colhead{$z=7.4$}&\colhead{$z=3.7$}&\colhead{$z=4.5$}&\colhead{$z=6.0$}&\colhead{$z=7.4$}}
\startdata
CCAT-p&4000&37&24&3&0.21&870&2100&17000&130000\\
CONCERTO&1200&\nodata&21&4&0.23&\nodata&1500&7500&59000\\
TIME&1000&\nodata&\nodata&1&0.12&\nodata&\nodata&6700&28000
\enddata
\tablecomments{The survey bandwidth assumed is $\Delta\nu=40$ GHz except at $z=7.4$, where $\Delta\nu = 28$ GHz. Values at $z=7.4$ may be adjusted to reflect the `optimistic' forecast simply by multiplying $\mathrm{S/N}$ by 3 and dividing survey times by 3.}
\end{deluxetable*}

We plot the expected sensitivities of CCAT-p, TIME, and CONCERTO in~\autoref{fig:pspec} given the $P(k)$ obtained in this work, \replaced{and report in~\autoref{tab:totSNR} the expected signal-to-noise ratio over all modes up to $k=1$ Mpc$^{-1}$. In both cases, we have accounted}{accounting} for the expected signal attenuation $W(k)$ due to beam smoothing\replaced{. W}{; w}e plot the noise in~\autoref{fig:pspec} as $\sigma_P(k)/W(k)$\replaced{, and calculate the signal-to-noise in~\autoref{tab:totSNR}}{. \autoref{tab:totSNR} reports the expected total signal-to-noise ratio over all modes up to $k=1$ Mpc$^{-1}$, calculated} as
\begin{equation}\mathrm{S/N}=\left[\sum_i\left(\frac{P(k_i)W(k_i)}{\sigma_P(k_i)}\right)^2\right]^{1/2},\end{equation}
summing over all $k$-bins with central values $k_i<1$ Mpc$^{-1}$.

Note the slope of the sensitivity curves at low $k$ for CCAT-p and CONCERTO. For the lowest redshifts, the signal is large enough for sample variance to be a significant if not dominant component of $\sigma_P(k)$. Therefore, at the low-$k$ end, where $P(k)\sim k^{-1}$ and $1/\sqrt{N_m(k)}\sim k^{-1}$, $\sigma_P(k)\sim k^{-2}$.\added{ (\autoref{fig:pspec_Pnonly} from~\autoref{sec:SNR_Pnonly} shows the sensitivity curves if only instrumental noise is considered.)} At higher redshifts, where instrumental noise dominates $\sigma_P$, $P_n\sim k^0$ but $1/\sqrt{N_m(k)}\sim k^{-1}$ still, so $\sigma_P(k)\sim k^{-1}$. A similar argument holds for TIME at the redshifts it observes, except $1/\sqrt{N_m(k)}\sim k^{-1/2}$ at low $k$ (see~\autoref{sec:Nmodes}).

More importantly, there is a discontinuity in the slope of all sensitivity curves, at $k\sim0.1$--0.3 Mpc$^{-1}$. This fact stems from the limited spectral resolution of all instruments, which significantly affects the growth of $N_m(k)$ beyond a specific $k$ as line-of-sight modes become inaccessible (again, see~\autoref{sec:Nmodes}). The effect is particularly severe for TIME's line-scan survey strategy, as it targets only one angular dimension.

Nonetheless, a detection of the [\ion{C}{2}] signal at $z\sim6$ as predicted here is within reach. TIME is at a disadvantage due to the relatively limited $N_m(k)$ it probes, but as it is expected to deploy first out of the three surveys, it will at minimum set bleeding-edge upper limits on the $z\sim6$ [\ion{C}{2}] auto spectrum, and an extended campaign with deeper mapping could yield a tentative detection. For instance, if deployed on a high-quality Atacama site, TIME would be capable of approximately twice the mapping speed it could achieve at the ARO, which translates to half the noise power spectrum after the same survey time (TIME collaboration, private communication). Such a deployment may be realised, for example, on the proposed Leighton Chajnantor Telescope (LCT), a refurbishment and move of the Caltech Submillimeter Observatory (CSO) from Mauna Kea to the Chajnantor plateau.\footnote{See the slide set from Sunil Golwala's Jan 2019 seminar given to the Caltech Far-IR Science Interest Group, available at time of writing at \url{https://fir-sig.ipac.caltech.edu/system/media_files/binaries/29/original/190115GolwalaLCTIRSIGWeb.pdf}.} Either in isolation or in combination with the ARO deployment, a 1000-hour survey with TIME at the LCT is extremely competitive against the other surveys presented here. 

To be more exact about necessary extensions to survey times for a detection of $P(k)$ at small $k$, \autoref{tab:totSNR} also shows how much time would be required to achieve a signal-to-noise of unity at a $k$-bin centred at $k=0.026$ Mpc$^{-1}$ of width $\Delta k=0.035$ Mpc$^{-1}$. (The corresponding total all-k $\mathrm{S/N}$ varies, and is especially lower for TIME due to its shallower sensitivity limit curve.) Optimisation of survey areas can also increase $t_\mathrm{pix}$ and improve sensitivity, but depending on the criteria the optimal survey areas at $z\sim6$ are too small for the instruments considered (see~\autoref{sec:optarea}).

\autoref{tab:totSNR} also shows that surveys would need to be unrealistically lengthy to detect the expected signal at $z\sim8$. This comes with the caveat from the end of~\autoref{sec:CIImean} that the predicted $P(k)$ at $z\sim8$ is likely too low. However, even if $P(k)$ here is too conservative by an order of magnitude, none of the surveys above would be sensitive enough to even reach a signal-to-noise ratio of 1 with their fiducial survey programmes, and all would require 5--10 times greater time on sky for an all-$k$ signal-to-noise of 2--4. TIME at ARO is more competitive at this redshift range than at $z\sim6$, with map noise expected to be several times better than either CCAT-p or CONCERTO. However, the line-scan strategy limits relative detectability of $P(k)$ for $k\gtrsim10^{-1}$ Mpc$^{-1}$ and thus total signal-to-noise across the scales considered here.

A second generation of [\ion{C}{2}] line-intensity surveys might attain a fully three-dimensional, wide-field detection (i.e.~over $\gtrsim 1$ deg$^2$), potentially even through a significant upgrade to an existing instrument or extension of an existing survey. Sensitivities must improve over the immediate generation by at least an order of magnitude, however. This would enable a more confident detection of our forecast signal at $z\sim6$ in addition to a tentative detection at $z\sim8$.

In view of this, we note the significant upgrade potential for EoR-Spec on CCAT-p. The Phase I instrument assumed here only occupies one-third of one instrument module, when in fact the overall Prime-Cam design (i.e., the first-light instrument design for CCAT-p) can accommodate up to seven instrument modules~\citep{PCam}. An upgraded EoR-Spec configuration---or a more aggressive first-light configuration---would use two fully occupied instrument modules for six times the field of view and six times the mapping speed. Note also that the seven-module Prime-Cam design does not entirely fill the CCAT-p telescope's field of view, which could accommodate as many as 19 modules~\citep{CCATp}. Such `Phase II' extensions of the EoR-Spec array by factors of 6--18 would provide the generational leap that may enable a detection of the [\ion{C}{2}] at $z\sim7.4$ (given the more optimistic version of our prediction).

TIME also has a clear path forward from what we present here. In addition to potential deployment at another telescope like the LCT, one way to increase sensitivity may be to make use of more space by moving from a grating spectrometer to a more compact architecture. The on-chip spectrometer technology to be deployed in SuperSpec~\citep{SuperSpec} provides a potential technology path. Once SuperSpec successfully demonstrates this technology in the field, future work should consider what may be possible with hundreds or thousands of SuperSpec-style pixels instead of the 16 considered here for TIME.

\section{Conclusions}
\label{sec:conclusions}

We have simulated the [\ion{C}{2}] signal that three near-future experiments will attempt to observe, and the results indicate promising prospects for [\ion{C}{2}] detections at $z\lesssim6$. If foregrounds like Galactic dust and lower-redshift emission in other lines can be overcome, these experiments promise to significantly improve our understanding of high-redshift galaxies from the end of reionization onwards.

While the signal will be weaker at $z\sim8$, the upcoming generation of [\ion{C}{2}] intensity mappers should still be able to set interesting limits on [\ion{C}{2}] in the epoch of reionization. Furthermore, their $z\lesssim6$ results will distinguish between the wide range of high-redshift [\ion{C}{2}] predictions that currently exist. There is also ample scope for further simulation studies, for example by deploying the \textsc{UniverseMachine} framework on higher-resolution simulations to better anticipate the effect of lower-mass halos on the signal. All such future theoretical and observational study will work in tandem to significantly narrow the model space in a way that guides the next generation of wide-field [\ion{C}{2}] surveyors.

\acknowledgements{We thank Guilaine Lagache and Matthieu Bethermin for communications regarding the survey and instrument parameters for CONCERTO; Gordon Stacey, Dominik Riechers, Michael Niemack, and other members of the CCAT-p science working group for detailed discussions about EoR-Spec; and members of the TIME collaboration including Lorenzo Moncelsi, C.~Matt Bradford, and Jonathan Hunacek for similar communications on TIME. We further thank Peter Behroozi for communications about the \textsc{UniverseMachine} EDR, as well as useful comments on this work; Sebastian Dumitru for communications about his work including revisions to~\cite{Dumitru18} in preparation; and Hamsa Padmanabhan for insightful communications on [\ion{C}{2}] forecasts, including her work while it was still in preparation. We would like to acknowledge the organizers and participants of the `Cosmological Signals from Cosmic Dawn to the Present' workshop held at the Aspen Center for Physics, which is supported by National Science Foundation grant PHY-1607611. Finally, we thank the anonymous referee for thoughtful comments and suggestions that improved the paper. This research made use of NASA's Astrophysics Data System Bibliographic Services.}
\software{\texttt{hmf}~\citep{hmf}; Matplotlib~\citep{matplotlib}; Astropy, a community-developed core Python package for astronomy~\citep{astropy}; WebPlotDigitizer (\url{https://automeris.io/WebPlotDigitizer}).}




\bibliographystyle{aasjournal}
\bibliography{references,morerefs} 

\begin{thebibliography}{}
\expandafter\ifx\csname natexlab\endcsname\relax\def\natexlab#1{#1}\fi
\providecommand{\url}[1]{\href{#1}{#1}}

\bibitem[{{Adam} {et~al.}(2014){Adam}, {Comis}, {Mac{\'{\i}}as-P{\'e}rez},
  {Adane}, {Ade}, {Andr{\'e}}, {Beelen}, {Belier}, {Beno{\^i}t}, {Bideaud},
  {Billot}, {Boudou}, {Bourrion}, {Calvo}, {Catalano}, {Coiffard}, {D'Addabbo},
  {D{\'e}sert}, {Doyle}, {Goupy}, {Kramer}, {Leclercq}, {Martino}, {Mauskopf},
  {Mayet}, {Monfardini}, {Pajot}, {Pascale}, {Perotto}, {Pointecouteau},
  {Ponthieu}, {Rev{\'e}ret}, {Rodriguez}, {Savini}, {Schuster}, {Sievers},
  {Tucker}, \& {Zylka}}]{NIKA}
{Adam}, R., {Comis}, B., {Mac{\'{\i}}as-P{\'e}rez}, J.~F., {et~al.} 2014, \aap,
  569, A66

\bibitem[{{Adam} {et~al.}(2018){Adam}, {Adane}, {Ade}, {Andr{\'e}},
  {Andrianasolo}, {Aussel}, {Beelen}, {Beno{\^i}t}, {Bideaud}, {Billot},
  {Bourrion}, {Bracco}, {Calvo}, {Catalano}, {Coiffard}, {Comis}, {De Petris},
  {D{\'e}sert}, {Doyle}, {Driessen}, {Evans}, {Goupy}, {Kramer}, {Lagache},
  {Leclercq}, {Leggeri}, {Lestrade}, {Mac{\'{\i}}as-P{\'e}rez}, {Mauskopf},
  {Mayet}, {Maury}, {Monfardini}, {Navarro}, {Pascale}, {Perotto}, {Pisano},
  {Ponthieu}, {Rev{\'e}ret}, {Rigby}, {Ritacco}, {Romero}, {Roussel}, {Ruppin},
  {Schuster}, {Sievers}, {Triqueneaux}, {Tucker}, \& {Zylka}}]{Adam18}
{Adam}, R., {Adane}, A., {Ade}, P.~A.~R., {et~al.} 2018, \aap, 609, A115

\bibitem[{{Astropy Collaboration} {et~al.}(2013){Astropy Collaboration},
  {Robitaille}, {Tollerud}, {Greenfield}, {Droettboom}, {Bray}, {Aldcroft},
  {Davis}, {Ginsburg}, {Price-Whelan}, {Kerzendorf}, {Conley}, {Crighton},
  {Barbary}, {Muna}, {Ferguson}, {Grollier}, {Parikh}, {Nair}, {Unther},
  {Deil}, {Woillez}, {Conseil}, {Kramer}, {Turner}, {Singer}, {Fox}, {Weaver},
  {Zabalza}, {Edwards}, {Azalee Bostroem}, {Burke}, {Casey}, {Crawford},
  {Dencheva}, {Ely}, {Jenness}, {Labrie}, {Lim}, {Pierfederici}, {Pontzen},
  {Ptak}, {Refsdal}, {Servillat}, \& {Streicher}}]{astropy}
{Astropy Collaboration}, {Robitaille}, T.~P., {Tollerud}, E.~J., {et~al.} 2013,
  \aap, 558, A33

\bibitem[{{Behroozi} {et~al.}(2019){Behroozi}, {Wechsler}, {Hearin}, \&
  {Conroy}}]{UniverseMachine}
{Behroozi}, P., {Wechsler}, R.~H., {Hearin}, A.~P., \& {Conroy}, C. 2019,
  \mnras, 488, 3143

\bibitem[{{Bradford} {et~al.}(2002){Bradford}, {Stacey}, {Swain}, {Nikola},
  {Bolatto}, {Jackson}, {Savage}, {Davidson}, \& {Ade}}]{SPIFI}
{Bradford}, C.~M., {Stacey}, G.~J., {Swain}, M.~R., {et~al.} 2002, \ao, 41,
  2561

\bibitem[{{Bradford} {et~al.}(2004){Bradford}, {Ade}, {Aguirre}, {Bock},
  {Dragovan}, {Duband}, {Earle}, {Glenn}, {Matsuhara}, {Naylor}, {Nguyen},
  {Yun}, \& {Zmuidzinas}}]{ZSpec}
{Bradford}, C.~M., {Ade}, P.~A.~R., {Aguirre}, J.~E., {et~al.} 2004, in
  \procspie, Vol. 5498, Millimeter and Submillimeter Detectors for Astronomy
  II, 257

\bibitem[{{Brauher} {et~al.}(2008){Brauher}, {Dale}, \& {Helou}}]{Brauher08}
{Brauher}, J.~R., {Dale}, D.~A., \& {Helou}, G. 2008, \apjs, 178, 280

\bibitem[{{Casey} {et~al.}(2014){Casey}, {Narayanan}, \& {Cooray}}]{Casey14}
{Casey}, C.~M., {Narayanan}, D., \& {Cooray}, A. 2014, \physrep, 541, 45

\bibitem[{{Choi} {et~al.}(2019){Choi}, {Austermann}, {Basu}, {Battaglia},
  {Bertoldi}, {Chung}, {Cothard}, {Duff}, {Duell}, {Gallardo}, {Gao}, {Herter},
  {Hubmayr}, {Niemack}, {Nikola}, {Riechers}, {Rossi}, {Stacey}, {Stevens},
  {Vavagiakis}, \& {Vissers}}]{Choi19}
{Choi}, S.~K., {Austermann}, J., {Basu}, K., {et~al.} 2019, arXiv e-prints,
  arXiv:1908.10451

\bibitem[{{Cousin} {et~al.}(2016){Cousin}, {Buat}, {Boissier}, {Bethermin},
  {Roehlly}, \& {G{\'e}nois}}]{Cousin16}
{Cousin}, M., {Buat}, V., {Boissier}, S., {et~al.} 2016, \aap, 589, A109

\bibitem[{{Cousin} {et~al.}(2015{\natexlab{a}}){Cousin}, {Lagache},
  {Bethermin}, {Blaizot}, \& {Guiderdoni}}]{Cousin15a}
{Cousin}, M., {Lagache}, G., {Bethermin}, M., {Blaizot}, J., \& {Guiderdoni},
  B. 2015{\natexlab{a}}, \aap, 575, A32

\bibitem[{{Cousin} {et~al.}(2015{\natexlab{b}}){Cousin}, {Lagache},
  {Bethermin}, \& {Guiderdoni}}]{Cousin15b}
{Cousin}, M., {Lagache}, G., {Bethermin}, M., \& {Guiderdoni}, B.
  2015{\natexlab{b}}, \aap, 575, A33

\bibitem[{{Crites} {et~al.}(2014){Crites}, {Bock}, {Bradford}, {Chang},
  {Cooray}, {Duband}, {Gong}, {Hailey-Dunsheath}, {Hunacek}, {Koch}, {Li},
  {O'Brient}, {Prouve}, {Shirokoff}, {Silva}, {Staniszewski}, {Uzgil}, \&
  {Zemcov}}]{Crites14}
{Crites}, A.~T., {Bock}, J.~J., {Bradford}, C.~M., {et~al.} 2014, in \procspie,
  Vol. 9153, Millimeter, Submillimeter, and Far-Infrared Detectors and
  Instrumentation for Astronomy VII, 91531W

\bibitem[{{De Looze} {et~al.}(2014){De Looze}, {Cormier}, {Lebouteiller},
  {Madden}, {Baes}, {Bendo}, {Boquien}, {Boselli}, {Clements}, {Cortese},
  {Cooray}, {Galametz}, {Galliano}, {Graci{\'a}-Carpio}, {Isaak}, {Karczewski},
  {Parkin}, {Pellegrini}, {R{\'e}my-Ruyer}, {Spinoglio}, {Smith}, \&
  {Sturm}}]{DeLooze14}
{De Looze}, I., {Cormier}, D., {Lebouteiller}, V., {et~al.} 2014, \aap, 568,
  A62

\bibitem[{{Dumitru} {et~al.}(2019){Dumitru}, {Kulkarni}, {Lagache}, \&
  {Haehnelt}}]{Dumitru18}
{Dumitru}, S., {Kulkarni}, G., {Lagache}, G., \& {Haehnelt}, M.~G. 2019,
  \mnras, 485, 3486

\bibitem[{{Ferland} {et~al.}(2017){Ferland}, {Chatzikos}, {Guzm{\'a}n},
  {Lykins}, {van Hoof}, {Williams}, {Abel}, {Badnell}, {Keenan}, {Porter}, \&
  {Stancil}}]{CLOUDY}
{Ferland}, G.~J., {Chatzikos}, M., {Guzm{\'a}n}, F., {et~al.} 2017, \rmxaa, 53,
  385

\bibitem[{{Finlator} {et~al.}(2017){Finlator}, {Prescott}, {Oppenheimer},
  {Dav{\'e}}, {Zackrisson}, {Livermore}, {Finkelstein}, {Thompson}, \&
  {Huang}}]{Finlator17}
{Finlator}, K., {Prescott}, M.~K.~M., {Oppenheimer}, B.~D., {et~al.} 2017,
  \mnras, 464, 1633

\bibitem[{{Gong} {et~al.}(2012){Gong}, {Cooray}, {Silva}, {Santos}, {Bock},
  {Bradford}, \& {Zemcov}}]{Gong12}
{Gong}, Y., {Cooray}, A., {Silva}, M., {et~al.} 2012, \apj, 745, 49

\bibitem[{{Hemmati} {et~al.}(2017){Hemmati}, {Yan}, {Diaz-Santos}, {Armus},
  {Capak}, {Faisst}, \& {Masters}}]{Hemmati17}
{Hemmati}, S., {Yan}, L., {Diaz-Santos}, T., {et~al.} 2017, \apj, 834, 36

\bibitem[{{Herrera-Camus} {et~al.}(2015){Herrera-Camus}, {Bolatto}, {Wolfire},
  {Smith}, {Croxall}, {Kennicutt}, {Calzetti}, {Helou}, {Walter}, {Leroy},
  {Draine}, {Brandl}, {Armus}, {Sandstrom}, {Dale}, {Aniano}, {Meidt},
  {Boquien}, {Hunt}, {Galametz}, {Tabatabaei}, {Murphy}, {Appleton}, {Roussel},
  {Engelbracht}, \& {Beirao}}]{HC15}
{Herrera-Camus}, R., {Bolatto}, A.~D., {Wolfire}, M.~G., {et~al.} 2015, \apj,
  800, 1

\bibitem[{Hunter(2007)}]{matplotlib}
Hunter, J.~D. 2007, Computing In Science \& Engineering, 9, 90

\bibitem[{{Klypin} {et~al.}(2016){Klypin}, {Yepes}, {Gottl{\"o}ber}, {Prada},
  \& {He{\ss}}}]{BolshoiPlanck1}
{Klypin}, A., {Yepes}, G., {Gottl{\"o}ber}, S., {Prada}, F., \& {He{\ss}}, S.
  2016, \mnras, 457, 4340

\bibitem[{{Kov{\'a}cs}(2008)}]{Kovacs08}
{Kov{\'a}cs}, A. 2008, in \procspie, Vol. 7020, Millimeter and Submillimeter
  Detectors and Instrumentation for Astronomy IV, 702007

\bibitem[{{Kovetz} {et~al.}(2017){Kovetz}, {Viero}, {Lidz}, {Newburgh},
  {Rahman}, {Switzer}, {Kamionkowski}, {Aguirre}, {Alvarez}, {Bock}, {Bond},
  {Bower}, {Bradford}, {Breysse}, {Bull}, {Chang}, {Cheng}, {Chung}, {Cleary},
  {Corray}, {Crites}, {Croft}, {Dor{\'e}}, {Eastwood}, {Ferrara}, {Fonseca},
  {Jacobs}, {Keating}, {Lagache}, {Lakhlani}, {Liu}, {Moodley}, {Murray},
  {P{\'e}nin}, {Popping}, {Pullen}, {Reichers}, {Saito}, {Saliwanchik},
  {Santos}, {Somerville}, {Stacey}, {Stein}, {Villaescusa-Navarro}, {Visbal},
  {Weltman}, {Wolz}, \& {Zemcov}}]{Kovetz17}
{Kovetz}, E.~D., {Viero}, M.~P., {Lidz}, A., {et~al.} 2017, ArXiv e-prints,
  arXiv:1709.09066

\bibitem[{{Lagache}(2018)}]{Lagache18IAU}
{Lagache}, G. 2018, in Proc.~IAU, Vol.~12, Symp.~S333, Peering towards Cosmic
  Dawn, ed. V.~{Jeli{\'c}} \& T.~{van der Hulst}, 228--233

\bibitem[{{Lagache} {et~al.}(2018){Lagache}, {Cousin}, \&
  {Chatzikos}}]{Lagache18}
{Lagache}, G., {Cousin}, M., \& {Chatzikos}, M. 2018, \aap, 609, A130

\bibitem[{{Li} {et~al.}(2016){Li}, {Wechsler}, {Devaraj}, \& {Church}}]{Li15}
{Li}, T.~Y., {Wechsler}, R.~H., {Devaraj}, K., \& {Church}, S.~E. 2016, \apj,
  817, 169

\bibitem[{{Murray} {et~al.}(2013){Murray}, {Power}, \& {Robotham}}]{hmf}
{Murray}, S.~G., {Power}, C., \& {Robotham}, A.~S.~G. 2013, Astronomy and
  Computing, 3, 23

\bibitem[{{Oberst} {et~al.}(2011){Oberst}, {Parshley}, {Nikola}, {Stacey},
  {L{\"o}hr}, {Lane}, {Stark}, \& {Kamenetzky}}]{Oberst11}
{Oberst}, T.~E., {Parshley}, S.~C., {Nikola}, T., {et~al.} 2011, \apj, 739, 100

\bibitem[{{Padmanabhan}(2019)}]{Padmanabhan18}
{Padmanabhan}, H. 2019, \mnras, 488, 3014

\bibitem[{{Planck Collaboration} {et~al.}(2016){Planck Collaboration}, {Ade},
  {Aghanim}, {Arnaud}, {Ashdown}, {Aumont}, {Baccigalupi}, {Banday},
  {Barreiro}, {Bartlett}, \& et~al.}]{Planck15}
{Planck Collaboration}, {Ade}, P.~A.~R., {Aghanim}, N., {et~al.} 2016, \aap,
  594, A13

\bibitem[{{Pullen} {et~al.}(2018){Pullen}, {Serra}, {Chang}, {Dor{\'e}}, \&
  {Ho}}]{Pullen18}
{Pullen}, A.~R., {Serra}, P., {Chang}, T.-C., {Dor{\'e}}, O., \& {Ho}, S. 2018,
  \mnras, 478, 1911

\bibitem[{{Rodr{\'{\i}}guez-Puebla} {et~al.}(2016){Rodr{\'{\i}}guez-Puebla},
  {Behroozi}, {Primack}, {Klypin}, {Lee}, \& {Hellinger}}]{BolshoiPlanck2}
{Rodr{\'{\i}}guez-Puebla}, A., {Behroozi}, P., {Primack}, J., {et~al.} 2016,
  \mnras, 462, 893

\bibitem[{{Serra} {et~al.}(2016){Serra}, {Dor{\'e}}, \& {Lagache}}]{Serra16}
{Serra}, P., {Dor{\'e}}, O., \& {Lagache}, G. 2016, \apj, 833, 153

\bibitem[{{Sheth} {et~al.}(2001){Sheth}, {Mo}, \& {Tormen}}]{SMT}
{Sheth}, R.~K., {Mo}, H.~J., \& {Tormen}, G. 2001, \mnras, 323, 1

\bibitem[{{Shirokoff} {et~al.}(2014){Shirokoff}, {Barry}, {Bradford},
  {Chattopadhyay}, {Day}, {Doyle}, {Hailey-Dunsheath}, {Hollister},
  {Kov{\'a}cs}, {Leduc}, {McKenney}, {Mauskopf}, {Nguyen}, {O'Brient}, {Padin},
  {Reck}, {Swenson}, {Tucker}, \& {Zmuidzinas}}]{SuperSpec}
{Shirokoff}, E., {Barry}, P.~S., {Bradford}, C.~M., {et~al.} 2014, Journal of
  Low Temperature Physics, 176, 657

\bibitem[{{Silva} {et~al.}(2015){Silva}, {Santos}, {Cooray}, \&
  {Gong}}]{Silva15}
{Silva}, M., {Santos}, M.~G., {Cooray}, A., \& {Gong}, Y. 2015, \apj, 806, 209

\bibitem[{{Spinoglio} {et~al.}(2012){Spinoglio}, {Dasyra}, {Franceschini},
  {Gruppioni}, {Valiante}, \& {Isaak}}]{Spinoglio12}
{Spinoglio}, L., {Dasyra}, K.~M., {Franceschini}, A., {et~al.} 2012, \apj, 745,
  171

\bibitem[{{Spinoglio} {et~al.}(2014){Spinoglio}, {Dasyra}, {Franceschini},
  {Gruppioni}, {Valiante}, \& {Isaak}}]{Spinoglio12E}
---. 2014, \apj, 791, 138

\bibitem[{{Stacey} {et~al.}(2018){Stacey}, {Aravena}, {Basu}, {Battaglia},
  {Beringue}, {Bertoldi}, {Bond}, {Breysse}, {Bustos}, {Chapman}, {Chung},
  {Cothard}, {Erler}, {Fich}, {Foreman}, {Gallardo}, {Giovanelli}, {Graf},
  {Haynes}, {Herrera-Camus}, {Herter}, {Hložek}, {Johnstone}, {Keating},
  {Magnelli}, {Meerburg}, {Meyers}, {Murray}, {Niemack}, {Nikola}, {Nolta},
  {Parshley}, {Riechers}, {Schilke}, {Scott}, {Stein}, {Stevens}, {Stutzki},
  {Vavagiakis}, \& {Viero}}]{CCATp}
{Stacey}, G.~J., {Aravena}, M., {Basu}, K., {et~al.} 2018, in \procspie, Vol.
  10700, Ground-based and Airborne Telescopes VII, 107001M.
\newblock \url{https://doi.org/10.1117/12.2314031}

\bibitem[{{Sun} {et~al.}(2018){Sun}, {Moncelsi}, {Viero}, {Silva}, {Bock},
  {Bradford}, {Chang}, {Cheng}, {Cooray}, {Crites}, {Hailey-Dunsheath},
  {Uzgil}, {Hunacek}, \& {Zemcov}}]{Sun16}
{Sun}, G., {Moncelsi}, L., {Viero}, M.~P., {et~al.} 2018, \apj, 856, 107

\bibitem[{{Uzgil} {et~al.}(2014){Uzgil}, {Aguirre}, {Bradford}, \&
  {Lidz}}]{Uzgil14}
{Uzgil}, B.~D., {Aguirre}, J.~E., {Bradford}, C.~M., \& {Lidz}, A. 2014, \apj,
  793, 116

\bibitem[{{Vallini} {et~al.}(2015){Vallini}, {Gallerani}, {Ferrara},
  {Pallottini}, \& {Yue}}]{Vallini15}
{Vallini}, L., {Gallerani}, S., {Ferrara}, A., {Pallottini}, A., \& {Yue}, B.
  2015, \apj, 813, 36

\bibitem[{{Vavagiakis} {et~al.}(2018){Vavagiakis}, {Ahmed}, {Ali}, {Basu},
  {Battaglia}, {Bertoldi}, {Bond}, {Bustos}, {Chapman}, {Chung}, {Coppi},
  {Cothard}, {Dicker}, {Duell}, {Duff}, {Erler}, {Fich}, {Galitzki},
  {Gallardo}, {Henderson}, {Herter}, {Hilton}, {Hubmayr}, {Irwin}, {Koopman},
  {McMahon}, {Murray}, {Niemack}, {Nikolas}, {Nolta}, {Orlowski-Scherer},
  {Parshley}, {Riechers}, {Rossi}, {Scott}, {Sierra}, {Silva-Feaver}, {Simon},
  {Stacey}, {Stevens}, {Ullom}, {Vissers}, {Walker}, {Wollack}, {Xu}, \&
  {Zhu}}]{PCam}
{Vavagiakis}, E.~M., {Ahmed}, Z., {Ali}, A., {et~al.} 2018, in \procspie, Vol.
  10708, Millimeter, Submillimeter, and Far-Infrared Detectors and
  Instrumentation for Astronomy IX, 107081U.
\newblock \url{https://doi.org/10.1117/12.2313868}

\bibitem[{Walker(2015)}]{Walker}
Walker, C.~K. 2015, Terahertz Astronomy (Boca Raton: CRC Press).
\newblock \url{https://www.taylorfrancis.com/books/9781466570436}

\bibitem[{{Wechsler} \& {Tinker}(2018)}]{Wechsler18}
{Wechsler}, R.~H., \& {Tinker}, J.~L. 2018, \araa, 56, 435

\bibitem[{{Yue} {et~al.}(2015){Yue}, {Ferrara}, {Pallottini}, {Gallerani}, \&
  {Vallini}}]{Yue15}
{Yue}, B., {Ferrara}, A., {Pallottini}, A., {Gallerani}, S., \& {Vallini}, L.
  2015, \mnras, 450, 3829

\bibitem[{{Zmuidzinas}(2003)}]{Zmuidzinas03}
{Zmuidzinas}, J. 2003, \ao, 42, 4989

\end{thebibliography}




\appendix

\section{The Number of Fourier Modes in a Given Wavenumber Bin}
\label{sec:Nmodes}
The analytic expression of $N_m(k)$ is different for TIME versus CCAT-p or CONCERTO, simply because of the line-scan nature of the survey. Normally, we discretise the Fourier space in cells of $(2\pi)^3/V_\mathrm{surv}$ and divide this into the volume of the Fourier shell corresponding to the range $(k,k+\Delta k)$:
\begin{equation}N_m(k)=\frac{1}{2}\frac{4\pi k^2\,\Delta k}{(2\pi)^3/V_\mathrm{surv}}=\frac{k^2\,\Delta k\,V_\mathrm{surv}}{4\pi^2},\end{equation}
where the factor of $1/2$ comes in from the fact that the Fourier transform is of all real numbers and thus only half of the modes in the full 3D Fourier shell are independent.

However, in the case of TIME, we effectively work in a 2D Fourier space, ignoring the shortest dimension. Thus, we only get a circular slice of this shell, of area $2\pi k\Delta k$, with a resolution of $(2\pi)^2$ divided by the comoving area $A_\mathrm{surv}$ of the survey:
\begin{equation}N_{m,\mathrm{2D}}(k)=\frac{1}{2}\frac{2\pi k\,\Delta k}{(2\pi)^2/A_\mathrm{surv}}=\frac{k\,\Delta k\,A_\mathrm{surv}}{4\pi}.\end{equation}

Furthermore, the limited frequency resolution of all experiments means that beyond a cutoff $k_{\delta_\nu}$ (given by $\pi$ divided by the comoving voxel length along the line of sight), $N_m$ will grow by one less power of $k$. Quantitatively speaking, the surface area of a spherical segment truncated at two parallel planes, one intersecting the centre of the sphere and one separated from it by $k_{\delta_\nu}$, is given by $2\pi k_{\delta_\nu}k$. So for $k>k_{\delta_\nu}$,
the total $V_\mathrm{shell}$ is twice this times $\Delta k$, or $4\pi k_{\delta_\nu}k\,\Delta k$. Thus, in the 3D case,
\begin{equation}N_m(k)=\frac{\min{(k,k_{\delta_\nu})}\,k\,\Delta k\,V_\mathrm{surv}}{4\pi^2}.\end{equation}
For TIME, the area of the $k$-shell is
\begin{equation}A_\mathrm{shell}\approx\begin{cases} 4\arcsin{(k_{\delta_\nu}/k)}\, k\,\Delta k&(k>k_{\delta_\nu})\\2\pi k\,\Delta k&(k<k_{\delta_\nu})\end{cases}.\end{equation}
Then
\begin{equation}N_{m,\mathrm{2D}}(k)=\begin{cases} (2\pi^2)^{-1}\arcsin{(k_{\delta_\nu}/k)}\, k\,\Delta k\,A_\mathrm{surv}&(k>k_{\delta_\nu})\\(4\pi)^{-1}k\,\Delta k\,A_\mathrm{surv}&(k<k_{\delta_\nu})\end{cases}.\end{equation}

\section{Details of Noise-equivalent Quantities}
\label{sec:CCAT-psens}
This section of the appendix considers in further detail the quantities related to the sensitivity per sky pixel $\sigma_\text{pix}$ for each experiment, which in turn informs the instrumental noise power spectrum $P_n$ in~\autoref{sec:sensest}. We first consider~\autoref{sec:nefdexp} the expressions that allow calculation of $\sigma_\text{pix}$ from lower-level noise-equivalent quantities. Then, after recapping relevant parameters for each experiment considered in the main text, we show in~\autoref{sec:nefdcalc} that we can reproduce the quantities claimed for each experiment with minimal assumptions.

\subsection{Basic Expressions}
\label{sec:nefdexp}
The noise-equivalent flux density (NEFD) is effectively the noise per beam, and is given by system efficiencies, instrumental bandwidth, telescope aperture, and the total noise-equivalent power (NEP). Equation A8 of~\cite{Gong12} gives a dimensionally incorrect expression for the NEFD, so we refer to Equation 7.41 from~\cite{Walker}, presented below with a minor change in notation:
\begin{equation}
\operatorname{NEFD} = \frac{\operatorname{NEP}}{\eta_c\eta_tA_ee^{-\tau_\nu A}\delta_\nu}.\label{eq:nefd}
\end{equation}
Here $A_e$ is the effective aperture of the telescope, while $\delta_\nu$ is the bandwidth of the observation (the spectrometer channel bandwidth in this case). The NEP itself depends on frequency and detector bandwidth, and in background-limited operation at the frequencies considered in this work, we adapt Equation 46 of~\cite{Zmuidzinas03}, which provides an approximate expression for background-limited power uncertainty if the Rayleigh-Jeans temperature $T_0$ of the background radiation is known:
\begin{equation}
    \sigma_P\approx\frac{kT_0}{\sqrt{\operatorname{BW}\cdot\tau}}\cdot\operatorname{BW},
\end{equation}
where $\operatorname{BW}_\text{det}$ is the bandwidth seen by the detector, and $\tau$ the integration time. By convention, this quantity is the NEP when we refer this uncertainty to a 1 Hz post-detection bandwidth, or $\tau=0.5$ seconds. Then we find
\begin{equation}
    \operatorname{NEP}\approx 0.62\text{ aW Hz}^{-1/2}\left(\frac{T_0}{\operatorname{K}}\right)\sqrt{\frac{\operatorname{BW}_\text{det}}{\operatorname{GHz}}}.
\end{equation}
Note that $\operatorname{BW}_\text{det}$ is not necessarily equal to the bandwidth used in~\autoref{eq:nefd}---a grating spectrometer or Fabry-Perot interferometer will only expose each detector to the bandwidth per channel (so that the NEFD scales as $\delta_\nu^{-1/2}$), but detectors behind a Fourier-transform spectrometer must see the entire instrumental bandwidth (so that the NEFD scales as $\delta_\nu^{-1}$).

We define the aperture efficiency $\eta_A\equiv A_e/(\pi D^2/4)$, where $D$ is the dish diameter, and then $\eta\equiv\eta_c\eta_t\eta_Ae^{-\tau_\nu A}$, to encapsulate the end-to-end optical coupling efficiency, including atmospheric attenuation.

The NEP is usually taken to be the power incident across the solid angle projected from the detector to the sky~\citep[see Equation 7.38 of][for instance]{Walker}. Therefore, to get the sensitivity per pixel $\sigma_\mathrm{pix}$ as noise-equivalent intensity (NEI), we divide the NEFD by the beam solid angle $\Omega_\mathrm{beam}=\mathrm{FWHM}^2\cdot\pi/(4\ln{2})$ (which the dependence of the NEFD on $A_e$ broadly cancels out):
\begin{equation}
\sigma_\mathrm{pix} = \frac{\operatorname{NEFD}}{\Omega_\mathrm{beam}} = \frac{\operatorname{NEP}}{\eta(\pi D^2/4)\delta_\nu\Omega_\mathrm{beam}}.\label{eq:nepapprox}
\end{equation}

This expression, together with~\autoref{eq:nepapprox}, allow us\added{ in principle} to re-derive the sensitivities considered in~\autoref{sec:sensest} from scratch assuming background-limited operation.

\subsection{Approximate Expected Noise-equivalent Quantities for All Experiments}
\label{sec:nefdcalc}
We now recap the basic design of each experiment considered in the main text, considering various trade-offs involved in each.
\begin{itemize}
\item TIME uses a grating spectrometer. The size of the grating limits the number of beams on sky, but all pixels integrate simultaneously on all frequencies, and each detector only sees the light corresponding to the width of the spectral bin.
\item CCAT-p uses a Fabry--Perot interferometer (FPI), which effectively acts as a narrow-band filter in front of the detectors. This means that the detectors only see the light corresponding to the width of the spectral bin, as in a grating spectrometer, but the FPI must scan through the different spectral bins. Therefore, at some point, there is a penalty proportional to $N_\text{chan}$ (the number of spectral bins) on the final map noise, as each detector spends only $\sim1/N_\text{chan}$ of its time integrating in a given spectral bin.
\item CONCERTO uses a Fourier-transform spectrometer (FTS), and specifically a Martin--Puplett interferometer (MPI). The relevant detail is that an interferometer in front of the camera scans through different time delays, so that each detector records an interferogram that can then be Fourier-transformed into a spectrum. As a result, each detector must see the entire bandwidth of the instrument rather than the bandwidth of the spectral bin. Unlike the FPI, despite the fact that there is a scan in time delay, the pixels do effectively integrate down on all frequency channels as the scan proceeds, so no $N_\text{chan}$ penalty applies.
\end{itemize}

For the purposes of this discussion, we consider all experiments only at 250 GHz, summarising the relevant parameters in the first several columns of~\autoref{tab:nefdapprox}. We assume a diffraction-limited beam for all experiments, with a FWHM of 50 arcsec for $D=6$ m and 25 arcsec for $D=12$ m at 250 GHz. (Note that this assumption is acceptable even for CCAT-p, as its beam FWHM approaches the diffraction limit at lower frequencies.)

Calculating the NEP and NEFD requires not only the instrument parameters, but also assumed values for the background Rayleigh-Jeans temperature $T_0$ and the end-to-end optical efficiency of the system $\eta$. We assume that for all experiments, a detector couples to 30\% of the source emission, so that $\eta=0.3$ for dual-polarisation pixels (as in CCAT-p and CONCERTO) and $\eta=0.15$ for single-polarisation spectrometers (which are what are counted up for TIME). We also assume that $T_0$ is given by 300 K times a system emissivity of 5\% for CCAT-p and CONCERTO, or 20\% for TIME (as it is at a significantly less dry site than the others, with much lower atmospheric transmission). Then $T_0=15$ K for CCAT-p and CONCERTO, and 60 K for TIME.

We show the resulting approximate NEP and NEFD, as well as $\sigma_\mathrm{pix}$ (the NEI), in the final several columns of~\autoref{tab:nefdapprox}. The results match the figures provided by each collaboration to within a factor of 2. Although these approximate calculations are highly dependent on our assumptions and ignore many complexities surrounding direct-detection spectrometer noise, they do show that all claimed figures for the sensitivity per pixel are facially at sensible ranges.\added{

We do caution that our assumptions are at times highly optimistic---in particular, an assumption of 5\% system emissivity represents an extreme best-case scenario for both CONCERTO and CCAT-p, but especially so for CONCERTO (given the optical configuration, antenna design, site altitude, and other considerations). Compare also to~\cite{Dumitru18}, where the 155 mJy s$^{1/2}$ NEFD figure is extrapolated from NIKA2 sensitivities demonstrated on sky with 2 mm precipitable water vapour and an observing elevation of 60 degrees---5\% emissivity is unachievable under such high water vapour content along the line of sight. However, as in the main text, we leave a re-examination of CONCERTO sensitivities for future work by others.

Furthermore, given the work that the individual collaborations have put into their own publicly and privately communicated sensitivity figures, which likely incorporate greater detail than we are able to, we do not replace the figures of~\autoref{tab:allparams} with the figures of~\autoref{tab:nefdapprox}; nor should we expect the two sets of figures to match perfectly, for similar reasons.}

\begin{deluxetable}{ccccccccccc}
\tabletypesize{\footnotesize}
\tablewidth{0.9\linewidth}
\tablecaption{\label{tab:nefdapprox} Instrumental parameters and approximate sensitivities for all experiments at 250 GHz.}
\tablehead{\colhead{Experiment}&\colhead{Spectrometer}&$N_\text{pix}$&$N_\text{spec,eff}$&\colhead{$D$}&\colhead{$\delta_\nu$}&\colhead{$\Delta\nu$}&\colhead{$\operatorname{BW}_\text{det}$ for NEP}&\colhead{approx.~NEP}&\colhead{approx.~NEFD}&\colhead{approx.~$\sigma_\mathrm{pix}$}\\&\colhead{architecture}&&&\colhead{(m)}&\colhead{(GHz)}&\colhead{(GHz)}&\colhead{(GHz)}&\colhead{(aW Hz$^{-1/2}$)}&\colhead{(mJy s$^{1/2}$)}&\colhead{(MJy sr$^{-1}$ s$^{1/2}$)}}
\startdata
CCAT-p&Fabry-Perot&1004&20&6&2.5&210&2.5&15&69&1.0\\
CONCERTO&Martin-Puplett&3000&3000&12&1.5&160&160&120&230&14\\
TIME&Grating&32&32&12&1.9&96&1.9&51&160&9.5
\enddata
\tablecomments{Calculations of approximate NEP, NEFD, and $
\sigma_\text{pix}$ assume a background brightness temperature of 15 K (60 K for TIME), corresponding to a ground-based 300 K antenna with 5\% system emissivity (20\% for TIME). The effective count of spectrometers $N_\text{spec,eff}$ integrating simultaneously across the full band is given by $N_\text{pix}$ divided by the number of channels in the band multiplied by the number of channels simultaneously observed at any given instant. For CONCERTO, we are considering a region of overlap between the 125--300 GHz and 200--360 GHz bands; $\Delta\nu$ differs between the two by 10\%, and we take the smaller $\Delta\nu$ here. The TIME parameters are for the high-frequency 230--326 GHz band. The actual numbers provided by the collaborations are $\sigma_\text{pix}=0.86$ MJy sr$^{-1}$ s$^{1/2}$ for CCAT-p, $\operatorname{NEFD}=155$ mJy s$^{1/2}$ (yielding $\sigma_\text{pix}=11$ MJy sr$^{-1}$ s$^{1/2}$) for CONCERTO, and $\sigma_\text{pix}=11$ MJy sr$^{-1}$ s$^{1/2}$ for TIME.}
\end{deluxetable}

\replaced{B}{In any case, b}ased on $\sigma_\mathrm{pix}$ alone, CONCERTO and TIME appear to be at a significant disadvantage relative to CCAT-p, respectively due to the increased NEP from the FTS architecture (which sets $\operatorname{BW}_\text{det}$ equal to $\Delta\nu$ rather than $\delta_\nu$) and the atmospheric conditions at ARO (reflected in our estimates in the higher background temperature). However, recall that CCAT-p uses a scanning spectrometer, with the FPI transmitting two colours at once to the dichroic TES array. Therefore, with $\approx100$ spectral bins in the entire band and only two bins seen by the spectrometer at any given time, there is a fifty-fold penalty incurred in the integration time per pixel. We encapsulate this in~\autoref{tab:nefdapprox} by borrowing a notation from~\cite{Padmanabhan18} and noting the effective count of spectrometers $N_\text{spec,eff}$ with full simultaneous frequency coverage across the entire instrument band, which is simply equal to $N_\mathrm{pix}$ for CONCERTO and TIME, but is $N_\mathrm{pix}/50$ for CCAT-p. Thus $N_\text{spec,eff}$ takes the place of $N_\mathrm{pix}$ in~\autoref{eq:tpix} for CCAT-p alone. Since the final survey sensitivity is given by $\sigma_\mathrm{pix}t_\mathrm{pix}^{-1/2}\propto\sigma_\mathrm{pix}N_\text{spec,eff}^{-1/2}/\Omega_\text{surv}$, the large spectrometer count of CONCERTO and the small survey area of TIME compensate for their relatively high $\sigma_\text{pix}$, which is how the final map noise ends up around the same order of magnitude for all experiments (as shown in~\autoref{tab:allparams} in the main text).

\section{Optimisation of Survey Area}
\label{sec:optarea}
We consider two possible ways to optimise survey area based on sensitivity requirements at a given $k$:
\begin{itemize}
\item Fixing all parameters (including survey time), we set $\Omega_\text{surv}$ to the value that sets the sample variance contribution to $\sigma_P$ equal to that of instrumental noise. This is the approach described in Appendix D of~\cite{Li15}.
\item Fixing only a desired signal-to-noise ratio at a given $k$-bin with width $\Delta k$, and fixing the total spectrometer bandwidth of the survey, we set $\Omega_\text{surv}$ to the value that minimises the $t_\text{surv}$ required. This is also described in Karoumpis et al.~in prep.
\end{itemize}

Given a nominal survey area $\Omega_\text{surv,nom}$ and time per pixel $t_\text{pix,nom}$, the first way requires us to set
\begin{equation}
P(k) = P_n = \frac{\sigma_\text{pix}^2}{t_\text{pix}}V_\text{vox} = \frac{\sigma_\text{pix}^2}{t_\text{pix,nom}}\frac{t_\text{pix,nom}}{t_\text{pix}}\alpha(z)\Omega_\text{pix}\delta_\nu = \frac{\sigma_\text{pix}^2}{t_\text{pix,nom}}\frac{\Omega_\text{surv}}{\Omega_\text{surv,nom}}\alpha(z)\Omega_\text{pix}\delta_\nu,
\end{equation}
where, using $R(z)$ to denote the comoving distance to redshift $z$, $\alpha(z)\equiv[cR(z)^2/H(z)](1+z)^2/\nu_\text{rest}$ is the conversion from data cube volumes (in units of solid angle times frequency) to comoving volumes at redshift $z$ (so for instance, $V_\text{vox}=\alpha(z)\Omega_\text{pix}\delta_\nu$).

Solving for $\Omega_\text{surv}$, we find
\begin{equation}
\Omega_\text{surv,opt1} = \frac{P(k)}{\alpha(z)}\cdot\frac{\Omega_\text{surv,nom}/\Omega_\text{pix}}{\sigma_\text{pix}^2t_\text{pix,nom}^{-1}\delta_\nu}.
\end{equation}

As in the calculation of $P_n$, note that since $t_\text{pix,nom}\propto\Omega_\text{pix}$ the end result $\Omega_\text{surv,opt1}$ is independent of map pixel size.

Meanwhile, the latter approach yields
\begin{equation}\Omega_\text{surv,opt2} = \left(\frac{4\pi P(k)}{\sigma_P(k)}\right)^2(\alpha(z)\,\Delta\nu_\text{surv}\,k^2\,\Delta k)^{-1}.\end{equation}
This method is useful as the optimal area is independent of the signal and of most instrumental details, but the instrumental details are necessary to derive the corresponding $t_\text{surv}$, which has no guarantee of being reasonable (especially for high $P/\sigma_P$).

Since the above uses $N_m(k)$ for the 3D case, the calculation cannot be repeated verbatim for the optimal survey extent $\theta_\text{surv}$ for a line-scan survey like TIME. If we define $\beta(z)\equiv\alpha(z)/R(z)$ so that $A_\text{surv}=\beta(z)\theta_\text{surv}\Delta\nu_\text{surv}$,
\begin{equation}\theta_\text{surv,opt2} = \pi \left(\frac{4P(k)}{\sigma_P(k)}\right)^2(\beta(z)\,\Delta\nu_\text{surv}\,k\,\Delta k)^{-1}.\end{equation}

\autoref{tab:optarea} shows optimal survey areas for all surveys. For all calculations, we assume $k=0.026$ Mpc$^{-1}$ (before the kink in $N_m(k)$) and $\Delta k=0.035$. As in the main text, we take $\Delta\nu_\text{surv}=40$ GHz except at $z=7.9$ where we assume 28 GHz.

It is important to compare the results to the instrumental field of view for each survey. Any optimal areas smaller than or even almost equal to these numbers should be considered unrealistic, as observing in stare modes is impractical at these frequencies for sensitive imaging~\citep{Kovacs08}.

\begin{deluxetable}{cccccccccc}
\tabletypesize{\footnotesize}
\tablewidth{0.9\linewidth}
\tablecaption{\label{tab:optarea}Optimal survey areas for detection of the [\ion{C}{2}] $P(k)$ in a $k$-bin centred at $k=0.026$ Mpc$^{-1}$ with width $\Delta k=0.035$, calculated using the equations given in~\autoref{sec:optarea}.}
\tablehead{
\colhead{Redshift}&\colhead{$P(k=0.026\text{ Mpc}^{-1})$}&\colhead{$R(z)$}&\colhead{$\alpha(z)$}&\multicolumn{3}{c}{$\Omega_\text{surv,opt1}$}&\multicolumn{2}{c}{$\Omega_\text{surv,opt2}$}&\colhead{$\theta_\text{surv,opt2}$}\\
\colhead{}&\colhead{}&\colhead{}&\colhead{}&\colhead{CCAT-p}&\colhead{CONCERTO}&\colhead{TIME}&\colhead{$P/\sigma_P=1$}&\colhead{$P/\sigma_P=3$}&\colhead{$P/\sigma_P=1$}\\
\colhead{}&\colhead{(Jy$^2$ sr$^{-2}$ Mpc$^3$)}&\colhead{(Mpc)}&\colhead{(Mpc$^3$ hr sr$^{-1}$)}&\multicolumn{2}{c}{(deg$^2$)}&\colhead{(arcmin$^2$)}&\multicolumn{2}{c}{(deg$^2$)}&\colhead{(arcmin)}
}
\startdata
3.7&$8.1\times10^{10}$&7100&$1.3\times10^{-4}$&5.7&\nodata&\nodata&1.2&11.&75.\\
4.5&$1.0\times10^{10}$&7700&$1.6\times10^{-4}$&2.0&0.81&\nodata&0.95&8.6&63.\\
6.0&$4.0\times10^{ 8}$&8500&$2.1\times10^{-4}$&0.20&0.098&$5.5\times0.5$&0.69&6.2&51.\\
7.9&$3.0\times10^{ 7}$&9000&$2.9\times10^{-4}$&0.027&0.021&$2.3\times0.5$&0.81&7.2&62.
\enddata
\tablecomments{The fields of view for CCAT-p, CONCERTO, and TIME are respectively $0.66\times0.66=0.44$ deg$^2$ per array (expandable to $2.0\times1.3=2.6$ deg$^2$ with two tubes of three arrays each), $15'$-diameter circular (0.049 deg$^2$), and $13.6\times0.5$ arcmin$^2$. The survey bandwidth is 40 GHz for all redshifts except $z=7.9$, where we assume a survey bandwidth of 28 GHz.}
\end{deluxetable}

\section{Radiometric Sensitivities of Surveys, Excluding Sample Variance}
\label{sec:SNR_Pnonly}
\begin{figure*}
\centering\includegraphics[width=0.72\linewidth]{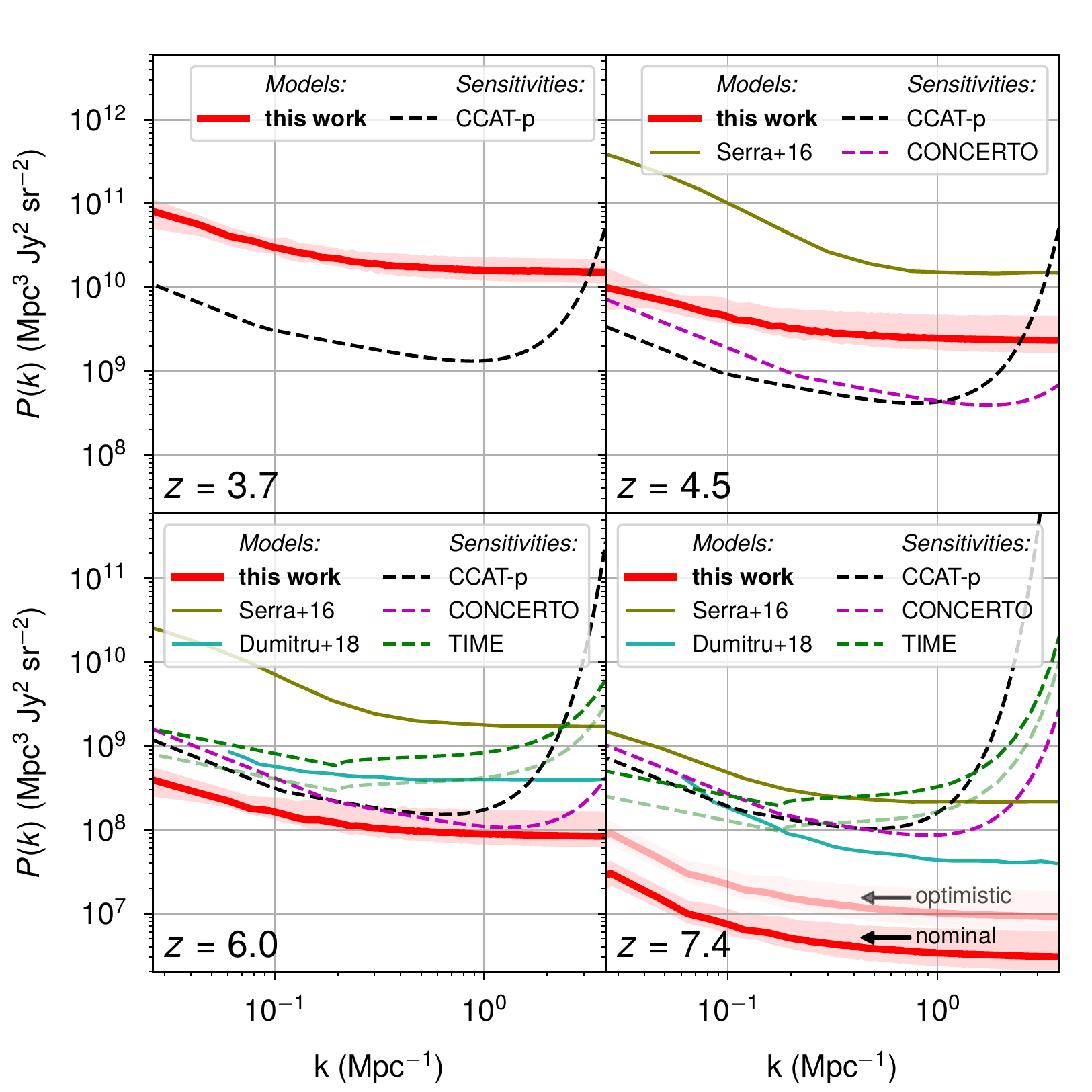}
\caption{Expected $1\sigma$ sensitivity limits (dashed curves) for CCAT-p, CONCERTO, and TIME (black, magenta, and green) given $k$-bins of width $\Delta k=0.035$ Mpc$^{-1}$, excluding sample variance and considering only thermal instrumental noise. We still plot $\sigma_P(k)/W(k)$ instead of just $\sigma_P(k)$; other details, including the $P(k)$ curves plotted, remain the same as in~\autoref{fig:pspec}.}
\label{fig:pspec_Pnonly}
\end{figure*}

The sensitivity curves of~\autoref{fig:pspec} account for both thermal instrumental noise and sample variance based on the fiducial model. While sample variance only significantly affects sensitivities for the lower two redshifts, this point may not be clear on a casual inspection of~\autoref{fig:pspec}. We therefore show in~\autoref{fig:pspec_Pnonly} a version of the same figure where the sensitivity curves exclude sample variance.
\end{document}